\documentclass[12pt,a4paper]{article}
\usepackage{epsfig}
\usepackage{axodraw}
\def\nl{\nonumber\\}
\def\beq{\begin{equation}}
\def\eeq{\end{equation}}
\def\beqar{\begin{eqnarray}}
\def\eeqar{\end{eqnarray}}
\def\bfi{\begin{figure}}
\def\efi{\end{figure}}
\def\btab{\begin{table}}
\def\etab{\end{table}}
\def\bce{\begin{center}}
\def\ece{\end{center}}
\def\bit{\begin{itemize}}
\def\eit{\end{itemize}}

\def\scrs{\scriptscriptstyle}
\def\text{\textstyle}

\def\al{\alpha}

\def\de{\delta}

\def\la{\lambda}
\def\si{\sigma}

\def\refeq#1{\mbox{(\ref{#1})}}
\def\reffi#1{\mbox{Fig.~\ref{#1}}}

\def\refse#1{\mbox{Sect.~\ref{#1}}}

\def\refapp#1{\mbox{Appendix~\ref{#1}}}
\def\citere#1{\mbox{Ref.~\cite{#1}}}
\def\citeres#1{\mbox{Refs.~\cite{#1}}}

\newcommand{\GeV}{\unskip\,\mathrm{GeV}}

\newcommand{\TeV}{\unskip\,\mathrm{TeV}}
\newcommand{\fba}{\unskip\,\mathrm{fb}}

\def\mathswitchr#1{\relax\ifmmode{\mathrm{#1}}\else$\mathrm{#1}$\fi}

\newcommand{\PV}{\mathswitch V}
\newcommand{\PW}{\mathswitchr W}

\def\mathswitch#1{\relax\ifmmode#1\else$#1$\fi}

\newcommand{\MV}{\mathswitch {M_\PV}}

\newcommand{\sw}{\mathswitch {s_{\scrs\PW}}}
\newcommand{\cw}{\mathswitch {c_{\scrs\PW}}}
\newcommand{\rw}{{\mathrm{W}}}

\def\ie{i.e.\ }
\def\eg{e.g.\ }

\def\wrt{wrt.\ }

\newcommand{\ord}{{\cal O}}

\newcommand{\ri}{\mathrm{i}}

\newcommand{\rR}{\mathrm{R}}
\newcommand{\rL}{\mathrm{L}}
\newcommand{\rT}{{\mathrm{T}}}
\newcommand{\rS}{{\mathrm{S}}}
\newcommand{\rd}{{\mathrm{d}}}
\newcommand{\rc}{{\mathrm{c}}}

\newcommand{\pT}{p_{\mathrm{T}}}
\newcommand{\pTcut}{p_{\mathrm{T}}^{\mathrm{cut}}}

\newcommand{\M}{{\cal {M}}}
\newcommand{\calL}{{\cal L}}

\newcommand{\NNLLa}{\stackrel{\mathrm{NNLL}}{=}}
\newcommand{\NLLa}{\stackrel{\mathrm{NLL}}{=}}
\newcommand{\NNLL}{\mathrm{NNLL}}
\newcommand{\NLL}{\mathrm{NLL}}

\newcommand{\LO}{\mathrm{LO}}
\newcommand{\NLO}{\mathrm{NLO}}
\newcommand{\NNLO}{\mathrm{NNLO}}
\newcommand{\shat}{{\hat s}}
\newcommand{\that}{{\hat t}}
\newcommand{\uhat}{{\hat u}}
\newcommand{\rhat}{{\hat r}}

\newcommand{\rar}{{\rightarrow}}

\newcommand{\ks}{k\hspace{-0.52em}/\hspace{0.02em}}

\newcommand{\ps}{p\hspace{-0.42em}/\hspace{-0.08em}}

\newcommand{\varepss}{\varepsilon \hspace{-0.48em}/\hspace{-0.05em}}

\newcommand{\MSBAR}{\overline{\mathrm{MS}}}
\newcommand{\msbar}{$\MSBAR$}
\newcommand{\qbar}{{\bar q}}
\newcommand{\A}{{\mathcal{A}}}
\newcommand{\F}{{\mathcal{F}}}
\newcommand{\smel}{\mathcal{S}}
\newcommand{\loops}{{J}}
\newcommand{\Mqq}{\M^{\qbar q}}
\newcommand{\OSa}{\stackrel{\mathrm{OS}}{=}}
\newcommand{\MSa}{\stackrel{\MSBAR}{=}}

\def\draftdate{\relax}
\def\mpar#1{\relax}
\def\mua{\relax}
\def\mda{\relax}
\def\mla{\relax}
\def\draft{
\def\thtystars{******************************}
\def\sixtystars{\thtystars\thtystars}
\typeout{}
\typeout{\sixtystars**}
\typeout{* Draft mode!
         For final version remove \protect\draft\space in source file *}
\typeout{\sixtystars**}
\typeout{}
\def\draftdate{\today}
\def\mua{\marginpar[\boldmath\hfil$\uparrow$]%
                   {\boldmath$\uparrow$\hfil}%
                    \typeout{marginpar: $\uparrow$}\ignorespaces}
\def\mda{\marginpar[\boldmath\hfil$\downarrow$]%
                   {\boldmath$\downarrow$\hfil}%
                    \typeout{marginpar: $\downarrow$}\ignorespaces}
\def\mla{\marginpar[\boldmath\hfil$\rightarrow$]%
                   {\boldmath$\leftarrow $\hfil}%
                    \typeout{marginpar: $\leftrightarrow$}\ignorespaces}
\def\Mua{\marginpar[\boldmath\hfil$\Uparrow$]%
                   {\boldmath$\Uparrow$\hfil}%
                    \typeout{marginpar: $\Uparrow$}\ignorespaces}
\def\Mda{\marginpar[\boldmath\hfil$\Downarrow$]%
                   {\boldmath$\Downarrow$\hfil}%
                    \typeout{marginpar: $\Downarrow$}\ignorespaces}
\def\Mla{\marginpar[\boldmath\hfil$\Rightarrow$]%
                   {\boldmath$\Leftarrow $\hfil}%
                    \typeout{marginpar: $\Leftrightarrow$}\ignorespaces}
\def\mpar##1{\marginpar{\hbadness10000%
                      \sloppy\hfuzz10pt\boldmath\bf##1}%
                      \typeout{marginpar: ##1}\ignorespaces}

\overfullrule 5pt
\oddsidemargin -15mm
\marginparwidth 29mm
}

\begin{document}

\thispagestyle{empty}
\def\thefootnote{\fnsymbol{footnote}}
\setcounter{footnote}{1}
\null
\draftdate
\hfill   TTP05-10\\
\strut\hfill  SFB/CPP-05-30\\
\strut\hfill hep-ph/0507178
\vskip 0cm
\vfill
\begin{center}
{\Large \bf
One-loop weak corrections to hadronic 
production of $Z$ bosons at large transverse momenta
\par}
\vskip 1em
{\large
{\sc Johann~H.~K\"uhn\footnote{Johann.Kuehn@physik.uni-karlsruhe.de}, 
A.~Kulesza\footnote{ania@particle.uni-karlsruhe.de},
S.~Pozzorini\footnote{pozzorin@particle.uni-karlsruhe.de},
M.~Schulze\footnote{schulze@particle.uni-karlsruhe.de} }}
\\[.5cm]
{\it Institut f\"ur Theoretische Teilchenphysik, 
Universit\"at Karlsruhe \\
D-76128 Karlsruhe, Germany}
\par
\end{center}\par
\vskip 1.0cm 
\vfill 
{\bf Abstract:} \par 
To match the precision of present and future measurements 
of $Z$-boson production at hadron colliders,
electroweak radiative corrections must be included
in the theory predictions.
In this paper we consider their effect on the 
transverse momentum ($p_\rT$) distribution of $Z$ bosons,
with emphasis on large $p_\rT$.
We evaluate, analytically and numerically, the full one-loop corrections 
for the parton scattering reaction $q\bar q \to Z g$
and its crossed variants.
In addition we derive compact approximate expressions which are valid 
in the high-energy region, where the weak corrections are strongly 
enhanced by logarithms of $\shat/M_W^2$.
These expressions include quadratic and single logarithms as well as 
those terms that are not logarithmically enhanced. 
This approximation, which  confirms and extends earlier results obtained to  
next-to-leading logarithmic accuracy, permits to reproduce the exact 
one-loop corrections with high precision.
Numerical results are presented for proton-proton
and proton-antiproton collisions.
The corrections are negative and their size increases with $p_\rT$.
For the Tevatron they amount up to $-7\%$ at 300 GeV.
For the LHC, where transverse momenta of 2 TeV or more can be reached,
corrections up to $-40\%$ are observed.
We also include the dominant two-loop effects of up to 8\% 
in our final LHC predictions.

\par
\vskip 1cm
\noindent
July 2005 
\par
\null
\setcounter{page}{0}
\clearpage
\def\thefootnote{\arabic{footnote}}
\setcounter{footnote}{0}

\newpage

\section{Introduction}

The study of gauge boson production has been among the primary 
goals of hadron colliders, starting with the discovery of the $W$ 
and $Z$ bosons more than two decades ago 
\cite{Arnison:1983rp}.
The investigation of the production dynamics, strictly predicted by 
the electroweak theory, constitutes one of the important tests
of the Standard Model. 
Furthermore, this reaction contributes to the background for 
many signals of new physics.
Being embedded in the
environment of hadronic collisions, the reaction necessarily
involves hadronic physics, like parton distributions, and
depends on the strong coupling constant. In turn, the cross
section for the production of $W$ and $Z$ bosons and their rapidity
distribution can be used to gauge the parton distribution
functions \cite{Catani:2000jh} which are important ingredients for the
prediction of numerous reaction rates. 

Differential distributions of gauge bosons, in rapidity as well
as in transverse momentum ($p_\rT$), 
have always been the subject of
theoretical and experimental studies.
In the region of small $p_\rT$, 
multiple gluon emission
plays an important role and contributions of arbitrary many
gluons must be resummed to arrive at a reliable prediction
\cite{ptres,ptres:tevlhc}.
At larger transverse momenta the final state of
the leading order process consists of a $W$ or $Z$ boson plus one
recoiling jet. QCD corrections in next-to-leading order
to this process are mandatory for the correct description 
and can amount to several tens of per cent depending on the observable under consideration, including jet definition, as well as the renormalization and factorization scales \cite{Ellis:1981hk}.
The evaluation of next-to-next-to-leading order
corrections involves two-loop virtual plus a variety of combined
virtual plus real corrections and is a topic presently pursued by
various groups (see \eg \citere{Gehrmann-DeRidder:2004xe}).

For the experiments at the Large Hadron Collider (LHC) a new
aspect comes into play. The high center-of-mass energy in
combination
with the enormous luminosity will allow to explore
parton-parton
scattering up to 
energies of several TeV 
and correspondingly production of gauge
bosons with transverse momenta up to \mbox{2 TeV} or even beyond. 
In this region electroweak
corrections from virtual weak boson exchange increase
strongly, with the dominant terms 
in $L$-loop approximation being 
leading logarithms of the form 
$\alpha^L\log^{2L}(\hat{s}/M_W^2)$,
next-to-leading logarithms of the form 
$\alpha^L\log^{2L-1}(\hat{s}/M_W^2)$, and so on.
These corrections, also known as electroweak Sudakov logarithms, may
well amount to several tens of percent.
They have been studied in
great detail for processes involving fermions in
\citeres{Kuhn:1999de,Pozzorini:2004rm}. 
Investigations on the
dominant and the next-to-leading logarithmic terms are also
available for reactions involving gauge and Higgs bosons
\cite{Denner:2001jv,Pozzorini:rs,Denner:2003wi}.
A recent survey of the literature on 
logarithmic electroweak corrections
can be found in \citere{Hollik:2004dz}.
The impact of these corrections
at hadron colliders has been studied in \citere{Accomando:2001fn}.
Specifically, hadronic $Z$-boson production at large $p_\rT$ has
been investigated in next-to-leading logarithmic 
approximation, including the two-loop terms \cite{Kuhn:2004em}. 
Numerical results for the complete one-loop terms 
have been presented in \citere{Maina:2004rb}.

It is the aim of this work to obtain an independent evaluation of
the complete one-loop weak 
corrections to the same
reaction, and to present the full result in analytic form. At
the partonic level
the reactions $q\bar{q}\rightarrow\mathrm{Z} g$, 
$qg\rightarrow\mathrm{Z} q$ and
$\bar{q}g\rightarrow\mathrm{Z} \bar{q}$ 
with $q=u,d,s,c$ or $b$ have to
be considered which are, however, trivially related by crossing
and appropriate exchange of coupling constants. We split the
corrections into an ''Abelian'' and a ''non-Abelian'' 
component. 
The ultraviolet divergences of the former are removed by the renormalization of the fermion wave functions as expected in Abelian theories.
The latter receives also divergent contributions from the renormalization 
of the $Z$-boson wave function and the electroweak parameters.
The electroweak coupling constants are renormalized 
in the framework of the modified minimal subtraction (\msbar) 
scheme. 
We present analytic results for the exact one-loop corrections
that permit to predict
separately the various quark-helicity contributions.
We also derive compact analytic expressions for the high-energy behaviour 
of the  corrections.
Here we include quadratic and linear logarithms as well as those 
terms that are not logarithmically enhanced at high energies but 
neglect all contributions of $\ord(M^2_W/\hat{s})$. 
The accuracy of this approximation is discussed in detail.

After convolution with parton distribution functions,
radiatively corrected predictions for transverse momentum distributions
of $Z$ bosons at hadron colliders are obtained.
Concerning perturbative QCD, these predictions are based on the lowest
order and thus proportional to $\alpha_\rS$.
To obtain realistic cross sections,
higher-order QCD corrections would have to be included, in 
next-to-leading or even next-to-next-to-leading order.

The paper is organized as follows: In \refse{se:kinematics} the Born
approximation, our conventions, the kinematics, and the parton
distributions are introduced. \refse{se:corr} is concerned with
a detailed description of the radiative corrections. 
In \refse{se:preliminaries} the strategy of our calculation 
is described and the
relevant Feynman diagrams are introduced. 
\refse{se:reduction}
is concerned with the algebraic reduction 
of the amplitudes to a set of basic Lorentz and Dirac structures that are multiplied by scalar form factors and scalar one-loop integrals.
Subsequently the renormalization is performed (\refse{se:renormalisation}) 
and a compact form for the results is given (\refse{se:results}), 
with a decomposition into Abelian and non-Abelian contributions.
Special attention is paid to the behaviour in the high-energy limit
(\refse{se:helimit}),
which can be cast into a compact form.
Indeed the quadratic and linear logarithms confirm the evaluation of the Sudakov logarithms obtained earlier in \citere{Kuhn:2004em}.
\refse{se:numerics} then contains a detailed discussion of numerical results,
both for $pp$ and $p\bar p$ collisions at 14 TeV and 2 TeV, respectively.
The leading and next-to-leading two-loop logarithmic terms 
are included in this numerical analysis.
We also discuss the sensitivity of the 
results towards the choice of the renormalization scheme and justify the use of the \msbar~scheme adopted in this paper.
\refse{se:conc} concludes with a brief summary.

\section{Conventions and kinematics}
\label{se:kinematics}

The  $\pT$ distribution of $Z$ bosons in the reaction 
$h_1 h_2 \to Z + \mathrm{jet}$ 
is given by 
\newcommand{\pdf}[4]{f_{#1,#2}(#3,#4)}
\beq
\label{hadroniccs}
\frac{\rd \si^{h_1 h_2}}{\rd \pT}=
\sum_{i,j}\int_0^1\rd x_1 \int_0^1\rd x_2
\;\theta(x_1 x_2-\hat\tau_{\rm min})
\pdf{h_1}{i}{x_1}{\mu^2}
\pdf{h_2}{j}{x_2}{\mu^2}
\frac{\rd \hat{\si}^{i j}}{\rd \pT}
,  
\eeq
where 
$\hat \tau_{\rm min} =(\pT+m_\rT)^2/s$,
$m_{\rT}=\sqrt{\pT^2+ M_Z^2}$  
and $\sqrt{s}$ is the collider energy.
The indices  $i, j$  denote initial state partons ($q, \bar q, g$) and
$f_{h_1,i}(x,\mu^2)$, $f_{h_2,j}(x,\mu^2)$
are the corresponding parton distribution functions.
$\hat {\si}^{ij}$ is the partonic cross section for 
the subprocess $i j \to Z k$ and the sum 
runs over all $i,j$ combinations corresponding to the subprocesses
\beq\label{processes}
\bar q q \to Z g,\quad
q\bar q\to Z g,\quad
g q \to Z  q,\quad
q g \to Z  q,\quad
\bar q g  \to Z  \bar q,\quad
g \bar q   \to Z  \bar q
.
\eeq
The Mandelstam variables for the subprocess $i j \to Z k$  are defined
in the standard way
\beq
\shat=(p_i+ p_j)^2 
,\qquad 
\that=(p_i- p_Z)^2 
,\qquad 
\uhat=(p_j-p_Z)^2
. 
\eeq
The momenta  $p_{i}$, $p_{j}$, $p_{k}$ of the partons are assumed to be massless, whereas 
$p_Z^2= M_Z^2$. 
In terms of $x_1,x_2,\pT$ and the collider energy $\sqrt{s}$ we have
\beq
\shat=x_1 x_2 s,\qquad
\that=\frac{ M_Z^2-\shat}{2}(1-\cos\theta),\qquad
\uhat=\frac{ M_Z^2-\shat}{2}(1+\cos\theta),
\eeq
with $\cos\theta=\sqrt{1- 4\pT^2 \shat/(\shat- M_Z^2)^2}$
corresponding to the cosine of the angle between the momenta $p_i$ and
$p_Z$ in the partonic center-of-mass frame.

The angular and the $\pT$ distribution 
for the unpolarized partonic subprocess $ij\to Z k$
read 
\beqar\label{partoniccs0}
\frac{\rd \hat{\si}^{i j}}{\rd \cos\theta}
&=&
\frac{\shat-M_Z^2}{32\pi N_{i j}\shat^2}
\;\overline{\sum}|\M^{i j}|^2
\eeqar
and
\beqar\label{partoniccs1}
\frac{\rd \hat{\si}^{i j}}{\rd \pT}
&=&
\frac{\pT}{8\pi N_{i j}\shat|\that-\uhat|}
\left[
\overline{\sum}|\M^{i j}|^2+(\that\leftrightarrow \uhat)
\right]
,
\eeqar
where
\beqar
\overline{\sum}=
\frac{1}{4}
\sum_{\mathrm{pol}} 
\sum_{\mathrm{col}} 
\eeqar
involves the sum over polarization and color as well as the average factor $1/4$ for initial-state polarization. 
The factor $1/N_{i j}$ in \refeq{partoniccs0}-\refeq{partoniccs1}, 
with 
$N_{\bar q q}=N_{q\bar q}=N_\rc^2$, 
$N_{g q}=N_{q g}=N_{\bar q g}=N_{g\bar q}=N_\rc (N_\rc^2-1)$, 
and $N_\rc=3$, accounts for  the initial-state colour average.
The factor 
\beq
\that-\uhat
=
(\shat-M_Z^2)\cos\theta
=
s \sqrt{(x_1 x_2 -\hat \tau_{\rm min})
(x_1 x_2 -\hat \tau_{\rm min}+{4p_\rT m_\rT}/{s})} 
\eeq
in the denominator of \refeq{partoniccs1}
gives rise to the so-called Jacobian peak,
which arises at $\cos\theta=0$ or, equivalently,
at $x_1 x_2=\hat \tau_{\rm min}$ and is 
smeared by the integration over the parton distribution functions.

The unpolarized squared matrix elements for the processes 
\refeq{processes} 
are related by the crossing-symmetry relations
\beq\label{crossing1}
\overline{\sum}|\M^{g q}|^2
= 
-\left.\overline{\sum}|\M^{\bar q q}|^2 
\right|_{\shat\leftrightarrow \that}
,\qquad
\overline{\sum}|\M^{\bar q g}|^2
= 
-\left.\overline{\sum}|\M^{\bar q q}|^2 
\right|_{\shat\leftrightarrow \uhat}
,
\eeq
and
\beq\label{crossing2}
\overline{\sum}|\M^{j i}|^2
= 
\left.\overline{\sum}|\M^{i j}|^2 
\right|_{\that\leftrightarrow \uhat}
.
\eeq
Moreover, as a result of CP symmetry,
\beqar
\overline{\sum}|\M^{\bar q q}|^2
= 
\left.\overline{\sum}|\M^{\bar q q}|^2 
\right|_{\that\leftrightarrow \uhat}
.
\eeqar
Using these symmetries we can write 
\beqar\label{hadroniccs2}
\lefteqn{
\frac{\rd \si^{h_1 h_2}}{\rd \pT}
=
\sum_{q=u,d,c,s,b}\int_0^1\rd x_1 \int_0^1\rd x_2
\;\theta(x_1 x_2-\hat\tau_{\rm min})
}\nl&&{}\times
\Biggl\{
\biggl[
\pdf{h_1}{q}{x_1}{\mu^2}\pdf{h_2}{\bar q}{x_2}{\mu^2}
+(1\leftrightarrow 2) \biggr]
\frac{\rd \hat{\si}^{\bar q q}}{\rd \pT}
\nl&&{}+
\biggl[
\biggl(
\pdf{h_1}{q}{x_1}{\mu^2}\pdf{h_2}{g}{x_2}{\mu^2}+
\pdf{h_1}{\bar q}{x_1}{\mu^2}\pdf{h_2}{g}{x_2}{\mu^2}
\biggr)
+(1\leftrightarrow 2) \biggr]
\frac{\rd \hat{\si}^{q g}}{\rd \pT}
\Biggr\}
.\nl
\eeqar
By means of \refeq{crossing1} and  \refeq{crossing2}
$\overline{\sum}|\M^{qg}|^2$ is
expressed in terms of $\overline{\sum}|\M^{\bar q q}|^2$,
\ie the explicit computation of the unpolarized squared matrix element
needs to be performed only for one of the six processes in \refeq{processes}. 

To lowest order in $\alpha$ and $\alpha_\rS$, for the $\bar q q\to Z g$ process
\cite{Kuhn:2004em} 
\beq\label{generalamplitude}
\overline{\sum}|\M_0^{\bar q q}|^2=
8 \pi^2 \alpha \alpha_\rS (N_\rc^2-1)
\sum_{\la=\rL,\rR} \left(I^Z_{q_\la}\right)^2 
\frac{\that^2+\uhat^2+2 M_Z^2 \shat}{\that\uhat} 
,
\eeq
where $\alpha=e^2/(4 \pi)$ and $\alpha_\rS=g_\rS^2/(4 \pi)$ 
are the electromagnetic and the strong coupling constants 
and $I^Z_{q_\la}$
represents the eigenvalue of the generator associated with the 
$Z$ boson in the representation corresponding to right-handed ($\la=\rR$) or
left-handed ($\la=\rL$) quarks $q_\la$.
In terms of the weak isospin $T^3_{q_\la}$ and the weak hypercharge $Y_{q_\la}$
\beq\label{couplfact0}
I^Z_{q_\la}=\frac{\cw}{\sw} T^3_{q_\la}-\frac{\sw}{\cw}\frac{Y_{q_\la}}{2},
\eeq
with the shorthands $\cw=\cos{\theta_\rw}$ and  $\sw=\sin{\theta_\rw}$
for the  weak mixing angle $\theta_\rw$. 

In general, for gauge couplings we adopt the conventions of \citere{Pozzorini:rs}.
With this notation the $g q\bar q$ vertex and the $V q\bar q$ vertices with $V=A,Z,W^\pm$ read 
\beqar
 \hspace{1cm}
\begin{picture}(30.,20.)(0,-3)
\ArrowLine(0,0)(-20,20)
\Text(-25,20)[rt]{$\bar q$}
\Gluon(0,0)(30,0){2.5}{4}
\Text(25,5)[b]{$G^\mu$}
\ArrowLine(-20,-20)(0,0)
\Text(-25,-20)[rb]{$q$}
\Vertex(0,0){1.5}
\end{picture}
&=& - \ri g_\rS  t^a \gamma^{\mu}
, \hspace{2cm}
\begin{picture}(30.,20.)(0,-3)
\ArrowLine(0,0)(-20,20)
\Text(-25,20)[rt]{$\bar q$}
\Photon(0,0)(30,0){2.5}{4}
\Text(25,5)[b]{$V^\mu$}
\ArrowLine(-20,-20)(0,0)
\Text(-25,-20)[rb]{$q'$}
\Vertex(0,0){1.5}
\end{picture}
=\ri e \gamma^{\mu}
\sum_{\lambda=\mathrm{R,L}} \omega_{\lambda}  
I^V_{q_\la {q_\la}^{\hspace{-1.5mm}'}},
\eeqar
where $\omega_{\la}$ are the chiral projectors
\beq\label{projectors}
\omega_{\rR}=\frac{1}{2}(1+\gamma_5)
,\qquad
\omega_{\rL}=\frac{1}{2}(1-\gamma_5),
\eeq
$t^a$ are the Gell-Mann matrices and $I^V$ are matrices in the weak isospin space.
For diagonal matrices such as $I^Z$ we write $I^Z_{q_\la {q_\la}^{\hspace{-1.5mm}'}}=
\de_{q q'} I^Z_{q_\la}$.
The triple gauge-bosons vertices read
\beqar
 \hspace{1cm}
\begin{picture}(30.,20.)(0,-3)
\Photon(0,0)(-20,20){2.5}{4}
\Text(-25,20)[rt]{$V_a^{\mu_1}$}
\Photon(0,0)(30,0){2.5}{4}
\Text(25,5)[b]{$V_c^{\mu_3}$}
\Photon(-20,-20)(0,0){2.5}{4}
\Text(-25,-20)[rb]{$V_b^{\mu_2}$}
\Vertex(0,0){1.5}
\end{picture}
&=& \frac{e}{\sw}\varepsilon^{V_a V_b V_c}
[
g^{\mu_1\mu_2}(k_1-k_2)^{\mu_3}
+g^{\mu_2\mu_3}(k_2-k_3)^{\mu_1}
\nl&&{}
+g^{\mu_3\mu_1}(k_3-k_1)^{\mu_2}
].
\eeqar
The definition of the antisymmetric tensor $\varepsilon$
as well as useful group-theoretical identities 
can be found in App.~B of \citere{Pozzorini:rs}.

\section{$\ord(\alpha)$ corrections}
\label{se:corr}
\newcommand{\diagtI}[1]{
\begin{picture}(94.,90.)(-47,-45)
\ArrowLine(0,20)(-40,20)
\Vertex(0,20){1.5}
\Photon(0,20)(40,20){2.5}{6}
\ArrowLine(-40,-20)(0,-20)
\Vertex(0,-20){1.5}
\Gluon(0,-20)(40,-20){-2.5}{6}
\ArrowLine(0,-20)(0,20)
\Text(0,-35)[t]{#1}
\end{picture}}

\newcommand{\diagtII}[1]{
\begin{picture}(94.,90.)(-47,-45)
\ArrowLine(0,20)(-40,20)
\Vertex(0,20){1.5}
\Gluon(0,20)(40,20){2.5}{6}
\ArrowLine(-40,-20)(0,-20)
\Vertex(0,-20){1.5}
\Photon(0,-20)(40,-20){-2.5}{6}
\ArrowLine(0,-20)(0,20)
\Text(0,-35)[t]{#1}
\end{picture}}

\newcommand{\diagcI}[1]{
\begin{picture}(94.,90.)(-47,-45)
\ArrowLine(0,20)(-40,20)
\Photon(0,20)(40,20){2.5}{6}
\ArrowLine(-40,-20)(0,-20)
\Vertex(0,-20){1.5}
\Gluon(0,-20)(40,-20){-2.5}{6}
\ArrowLine(0,-20)(0,0)\ArrowLine(0,0)(0,20)
\Text(0,-35)[t]{#1}
\BCirc(0,0){4}\Line(-2.83,-2.83)(2.83,2.83)\Line(-2.83,2.83)(2.83,-2.83)
\end{picture}}

\newcommand{\diagcII}[1]{
\begin{picture}(94.,90.)(-47,-45)
\ArrowLine(0,20)(-40,20)
\Vertex(0,20){1.5}
\Gluon(0,20)(40,20){2.5}{6}
\ArrowLine(-40,-20)(0,-20)
\Vertex(0,-20){1.5}
\Photon(0,-20)(40,-20){-2.5}{6}
\ArrowLine(0,-20)(0,0)\ArrowLine(0,0)(0,20)
\Text(0,-35)[t]{#1}
\BCirc(0,0){4}\Line(-2.83,-2.83)(2.83,2.83)\Line(-2.83,2.83)(2.83,-2.83)
\end{picture}}

\newcommand{\diagcIII}[1]{
\begin{picture}(94.,90.)(-47,-45)
\ArrowLine(0,20)(-40,20)
\Photon(0,20)(40,20){2.5}{6}
\ArrowLine(-40,-20)(0,-20)
\Vertex(0,-20){1.5}
\Gluon(0,-20)(40,-20){-2.5}{6}
\ArrowLine(0,-20)(0,20)
\Text(0,-35)[t]{#1}
\BCirc(0,-20){4}\Line(-2.83,-22.83)(2.83,-17.18)\Line(-2.83,-17.18)(2.83,-22.83)
\end{picture}}

\newcommand{\diagcIV}[1]{
\begin{picture}(94.,90.)(-47,-45)
\ArrowLine(0,20)(-40,20)
\Vertex(0,20){1.5}
\Gluon(0,20)(40,20){2.5}{6}
\ArrowLine(-40,-20)(0,-20)
\Vertex(0,-20){1.5}
\Photon(0,-20)(40,-20){-2.5}{6}
\ArrowLine(0,-20)(0,20)
\Text(0,-35)[t]{#1}
\BCirc(0,20){4}\Line(-2.82,17.18)(2.83,22.82)\Line(-2.83,22.83)(2.83,17.18)
\end{picture}}

\newcommand{\diagcV}[1]{
\begin{picture}(94.,90.)(-47,-45)
\ArrowLine(0,20)(-40,20)
\Photon(0,20)(40,20){2.5}{6}
\ArrowLine(-40,-20)(0,-20)
\Vertex(0,-20){1.5}
\Gluon(0,-20)(40,-20){-2.5}{6}
\ArrowLine(0,-20)(0,20)
\Text(0,-35)[t]{#1}
\BCirc(0,20){4}\Line(-2.82,17.18)(2.83,22.82)\Line(-2.83,22.83)(2.83,17.18)
\end{picture}}

\newcommand{\diagcVI}[1]{
\begin{picture}(94.,90.)(-47,-45)
\ArrowLine(0,20)(-40,20)
\Vertex(0,20){1.5}
\Gluon(0,20)(40,20){2.5}{6}
\ArrowLine(-40,-20)(0,-20)
\Vertex(0,-20){1.5}
\Photon(0,-20)(40,-20){-2.5}{6}
\ArrowLine(0,-20)(0,20)
\Text(0,-35)[t]{#1}
\BCirc(0,-20){4}\Line(-2.83,-22.83)(2.83,-17.18)\Line(-2.83,-17.18)(2.83,-22.83)
\end{picture}}

\newcommand{\diagsI}[1]{
\begin{picture}(94.,90.)(-47,-45)
\ArrowLine(0,20)(-40,20)
\Vertex(0,20){1.5}
\Photon(0,20)(40,20){2.5}{6}
\ArrowLine(-40,-20)(0,-20)
\Vertex(0,-20){1.5}
\Gluon(0,-20)(40,-20){-2.5}{6}
\PhotonArc(0,0)(10,90,270){-2.5}{4}\Text(-15,0)[r]{$\scriptstyle{V}$}
\ArrowLine(0,-20)(0,-10)\Vertex(0,-10){1.5}\ArrowLine(0,-10)(0,10)\Vertex(0,10){1.5}\ArrowLine(0,10)(0,20)
\Text(0,-35)[t]{#1}
\end{picture}}

\newcommand{\diagsII}[1]{
\begin{picture}(94.,90.)(-47,-45)
\ArrowLine(0,20)(-40,20)
\Vertex(0,20){1.5}
\Gluon(0,20)(40,20){2.5}{6}
\ArrowLine(-40,-20)(0,-20)
\Vertex(0,-20){1.5}
\Photon(0,-20)(40,-20){-2.5}{6}
\PhotonArc(0,0)(10,90,270){-2.5}{4}\Text(-15,0)[r]{$\scriptstyle{V}$}
\ArrowLine(0,-20)(0,-10)\Vertex(0,-10){1.5}\ArrowLine(0,-10)(0,10)\Vertex(0,10){1.5}\ArrowLine(0,10)(0,20)
\Text(0,-35)[t]{#1}
\end{picture}}

\newcommand{\diagvI}[1]{
\begin{picture}(94.,90.)(-47,-45)
\ArrowLine(0,20)(-40,20)
\Vertex(0,20){1.5}
\Photon(0,20)(40,20){2.5}{6}
\ArrowLine(-40,-20)(-15,-20)
\Vertex(-15,-20){1.5}
\Vertex(15,-20){1.5}
\Gluon(15,-20)(40,-20){-2.5}{3.5}
\ArrowLine(-15,-20)(15,-20)
\Photon(-15,-20)(0,5){2.5}{4.5}\Text(-13,-5)[r]{$\scriptstyle{V}$}\ArrowLine(15,-20)(0,5)
\Vertex(0,5){1.5}\ArrowLine(0,5)(0,20)
\Text(0,-35)[t]{#1}
\end{picture}}

\newcommand{\diagvII}[1]{
\begin{picture}(94.,90.)(-47,-45)
\ArrowLine(-15,20)(-40,20)
\Vertex(-15,20){1.5}
\Vertex(15,20){1.5}
\Gluon(15,20)(40,20){2.5}{3.5}
\ArrowLine(15,20)(-15,20)
\Photon(0,-5)(-15,20){2.5}{4.5}\Text(-13,5)[r]{$\scriptstyle{V}$}\ArrowLine(0,-5)(15,20)
\Vertex(0,-5){1.5}\ArrowLine(0,-20)(0,-5)
\ArrowLine(-40,-20)(0,-20)
\Vertex(0,-20){1.5}
\Photon(0,-20)(40,-20){-2.5}{6}
\Text(0,-35)[t]{#1}
\end{picture}}

\newcommand{\diagvIII}[1]{
\begin{picture}(94.,90.)(-47,-45)
\ArrowLine(-15,20)(-40,20)
\Vertex(-15,20){1.5}
\Vertex(15,20){1.5}
\Photon(15,20)(40,20){2.5}{3.5}
\ArrowLine(15,20)(-15,20)
\Photon(0,-5)(-15,20){2.5}{4.5}\Text(-13,5)[r]{$\scriptstyle{V}$}\ArrowLine(0,-5)(15,20)
\Vertex(0,-5){1.5}\ArrowLine(0,-20)(0,-5)
\ArrowLine(-40,-20)(0,-20)
\Vertex(0,-20){1.5}
\Gluon(0,-20)(40,-20){-2.5}{6}
\Text(0,-35)[t]{#1}
\end{picture}}

\newcommand{\diagvIV}[1]{
\begin{picture}(94.,90.)(-47,-45)
\ArrowLine(0,20)(-40,20)
\Vertex(0,20){1.5}
\Gluon(0,20)(40,20){2.5}{6}
\ArrowLine(-40,-20)(-15,-20)
\Vertex(-15,-20){1.5}
\Vertex(15,-20){1.5}
\Photon(15,-20)(40,-20){-2.5}{3.5}
\ArrowLine(-15,-20)(15,-20)
\Photon(-15,-20)(0,5){2.5}{4.5}\Text(-13,-5)[r]{$\scriptstyle{V}$}\ArrowLine(15,-20)(0,5)
\Vertex(0,5){1.5}\ArrowLine(0,5)(0,20)
\Text(0,-35)[t]{#1}
\end{picture}}

\newcommand{\diagvV}[1]{
\begin{picture}(94.,90.)(-47,-45)
\ArrowLine(-15,20)(-40,20)
\Vertex(-15,20){1.5}
\Vertex(15,20){1.5}
\Photon(15,20)(40,20){2.5}{3.5}
\Photon(15,20)(-15,20){-2.5}{4.5}\Text(0,25)[b]{$\scriptstyle{W^\pm}$}
\ArrowLine(0,-5)(-15,20)\Photon(0,-5)(15,20){2.5}{4.5}\Text(13,5)[l]{$\scriptstyle{W^\mp}$}
\Vertex(0,-5){1.5}\ArrowLine(0,-20)(0,-5)
\ArrowLine(-40,-20)(0,-20)
\Vertex(0,-20){1.5}
\Gluon(0,-20)(40,-20){-2.5}{6}
\Text(0,-35)[t]{#1}
\end{picture}}

\newcommand{\diagvVI}[1]{
\begin{picture}(94.,90.)(-47,-45)
\ArrowLine(0,20)(-40,20)
\Vertex(0,20){1.5}
\Gluon(0,20)(40,20){2.5}{6}
\ArrowLine(-40,-20)(-15,-20)
\Vertex(-15,-20){1.5}
\Vertex(15,-20){1.5}
\Photon(15,-20)(40,-20){-2.5}{3.5}
\Photon(-15,-20)(15,-20){-2.5}{4.5}\Text(0,-25)[t]{$\scriptstyle{W^\pm}$}
\ArrowLine(-15,-20)(0,5)\Photon(15,-20)(0,5){2.5}{4.5}\Text(13,-5)[l]{$\scriptstyle{W^\mp}$}
\Vertex(0,5){1.5}\ArrowLine(0,5)(0,20)
\Text(0,-35)[t]{#1}
\end{picture}}

\newcommand{\diagbI}[1]{
\begin{picture}(94.,90.)(-47,-45)
\ArrowLine(-15,20)(-40,20)
\Vertex(-15,20){1.5}
\Vertex(15,20){1.5}
\Photon(15,20)(40,20){2.5}{3.5}
\ArrowLine(15,20)(-15,20)
\ArrowLine(-40,-20)(-15,-20)
\Vertex(-15,-20){1.5}
\Vertex(15,-20){1.5}
\Gluon(15,-20)(40,-20){-2.5}{3.5}
\ArrowLine(-15,-20)(15,-20)
\Photon(-15,-20)(-15,20){2.5}{5}\Text(-20,0)[r]{$\scriptstyle{V}$}
\ArrowLine(15,-20)(15,20)
\Text(0,-35)[t]{#1}
\end{picture}}

\newcommand{\diagbII}[1]{
\begin{picture}(94.,90.)(-47,-45)
\ArrowLine(-15,20)(-40,20)
\Vertex(-15,20){1.5}
\Vertex(15,20){1.5}
\Gluon(15,20)(40,20){2.5}{3.5}
\ArrowLine(15,20)(-15,20)
\ArrowLine(-40,-20)(-15,-20)
\Vertex(-15,-20){1.5}
\Vertex(15,-20){1.5}
\Photon(15,-20)(40,-20){-2.5}{3.5}
\ArrowLine(-15,-20)(15,-20)
\Photon(-15,-20)(-15,20){2.5}{5}\Text(-20,0)[r]{$\scriptstyle{V}$}
\ArrowLine(15,-20)(15,20)
\Text(0,-35)[t]{#1}
\end{picture}}

\newcommand{\diagbIII}[1]{
\begin{picture}(94.,90.)(-47,-45)
\ArrowLine(-15,20)(-40,20)
\Vertex(-15,20){1.5}
\Vertex(15,20){1.5}
\Photon(15,20)(40,20){2.5}{3.5}
\Photon(15,20)(-15,20){-2.5}{4.5}\Text(0,25)[b]{$\scriptstyle{W^\pm}$}
\ArrowLine(-40,-20)(-15,-20)
\Vertex(-15,-20){1.5}
\Vertex(15,-20){1.5}
\Gluon(15,-20)(40,-20){-2.5}{3.5}
\ArrowLine(-15,-20)(15,-20)
\Photon(-15,-20)(15,20){2.5}{7}\Text(13,0)[lb]{$\scriptstyle{W^\mp}$}
\ArrowLine(15,-20)(-15,20)
\Text(0,-35)[t]{#1}
\end{picture}}

\newcommand{\diagqqzct}{
\begin{picture}(40.,20.)(0,-3)
\ArrowLine(0,0)(-20,20)
\Photon(0,0)(30,0){2.5}{4}
\ArrowLine(-20,-20)(0,0)
\BCirc(0,0){4}\Line(-2.83,-2.83)(2.83,2.83)\Line(-2.83,2.83)(2.83,-2.83)
\end{picture}}

\newcommand{\diagqqgct}{
\begin{picture}(40.,20.)(0,-3)
\ArrowLine(0,0)(-20,20)
\Gluon(0,0)(30,0){2.5}{4}
\ArrowLine(-20,-20)(0,0)
\BCirc(0,0){4}\Line(-2.83,-2.83)(2.83,2.83)\Line(-2.83,2.83)(2.83,-2.83)
\end{picture}}

\newcommand{\diagqqct}{
\begin{picture}(40.,10.)(0,-3)
\ArrowLine(0,0)(-30,0)
\ArrowLine(30,0)(0,0)
\BCirc(0,0){4}\Line(-2.83,-2.83)(2.83,2.83)\Line(-2.83,2.83)(2.83,-2.83)
\end{picture}}

In this section we present the one-loop weak 
corrections to the process $\qbar q \to Z g$.
The algebraic reduction 
to standard matrix elements and scalar integrals   
is described in \refse{se:reduction}.
Renormalization, in the \msbar~and the on-shell scheme, 
is discussed in \refse{se:renormalisation}.
In \refse{se:results} we present analytical results for the unpolarized squared matrix elements, and the high-energy behaviour of the corrections is derived in \refse{se:helimit}.

\subsection{Preliminaries}
\label{se:preliminaries}
As discussed in the previous section, the 6 different processes relevant 
for $Z+1$ jet production are related by crossing symmetries. 
It is thus sufficient to consider only one of these processes.
{\unitlength 1pt \small
\begin{figure}
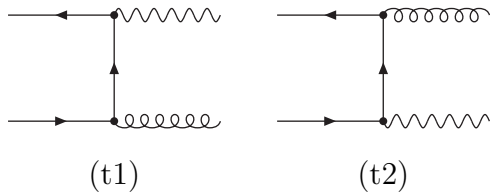

\begin{center}
\diagtI{(t1)}
\diagtII{(t2)}
\end{center}
 \caption{Tree-level Feynman diagrams for the process $\bar q q \to Z g$.}
 \label{fig:treediags}
 \end{figure}}
In the following we derive the one-loop corrections for the $\qbar q\to Z g$ process.
The matrix element 
\beqar\label{oneloopsplitting}
\Mqq_1&=&\Mqq_0+\delta \Mqq_{1}
\eeqar
is expressed as a function of the Mandelstam invariants 
\beq
\shat=(p_{\bar q}+ p_q)^2 
,\qquad 
\that=(p_{\bar q}- p_Z)^2 
,\qquad 
\uhat=(p_q-p_Z)^2.
\eeq
The Born contribution, $\Mqq_0$, results from the $t$- and $u$-channel diagrams 
of \reffi{fig:treediags}.
The loop and counterterm (CT) 
diagrams contributing to the corrections, 
\beqar\label{CTsplitting}
\delta \Mqq_1&=&
\delta \Mqq_{1,\mathrm{loops}}+
\delta \Mqq_{1,\mathrm{CT}},
\eeqar
are depicted in \reffi{fig:loopdiags} and \reffi{fig:ctdiags}, respectively.
{\unitlength 1pt \small
\begin{figure}
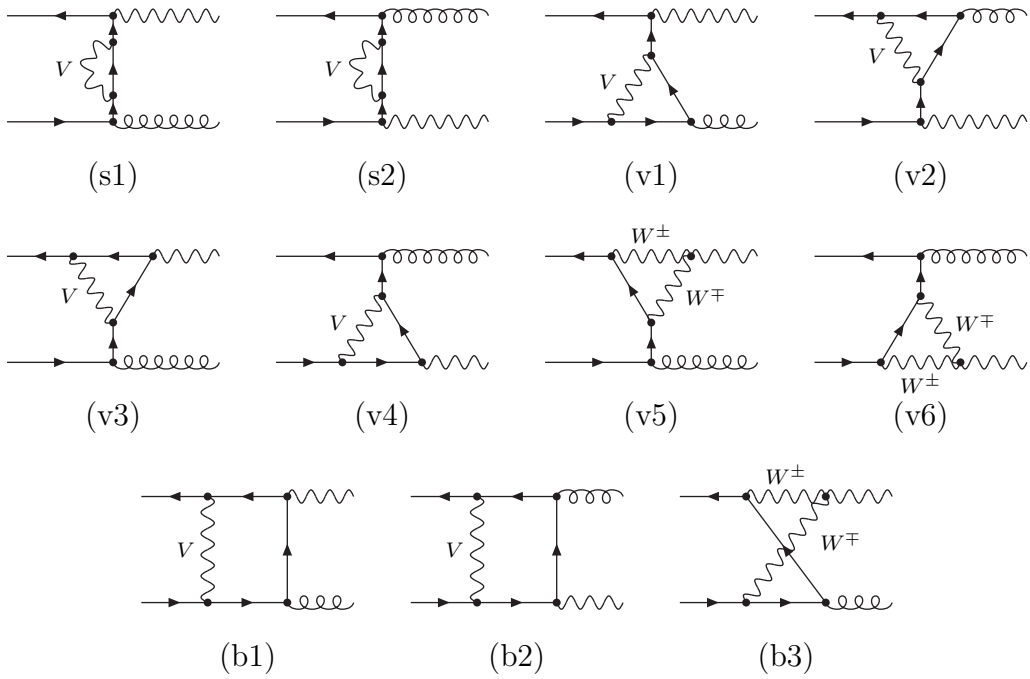

\begin{center}
\diagsI{(s1)}
\diagsII{(s2)}
\diagvI{(v1)}
\diagvII{(v2)}
\diagvIII{(v3)}
\diagvIV{(v4)}
\diagvV{(v5)}
\diagvVI{(v6)}
\diagbI{(b1)}
\diagbII{(b2)}
\diagbIII{(b3)}
\end{center}
 \caption{
One-loop Feynman diagrams for the process $\bar q q \to Z g$.
The diagrams  v5, v6 and b3 involve only 
charged weak bosons, $W^\pm$, whereas the other diagrams receive contributions from neutral and charged 
weak bosons, $V=Z,W^\pm$.
}
 \label{fig:loopdiags}
 \end{figure}}
{\unitlength 1pt \small
\begin{figure}
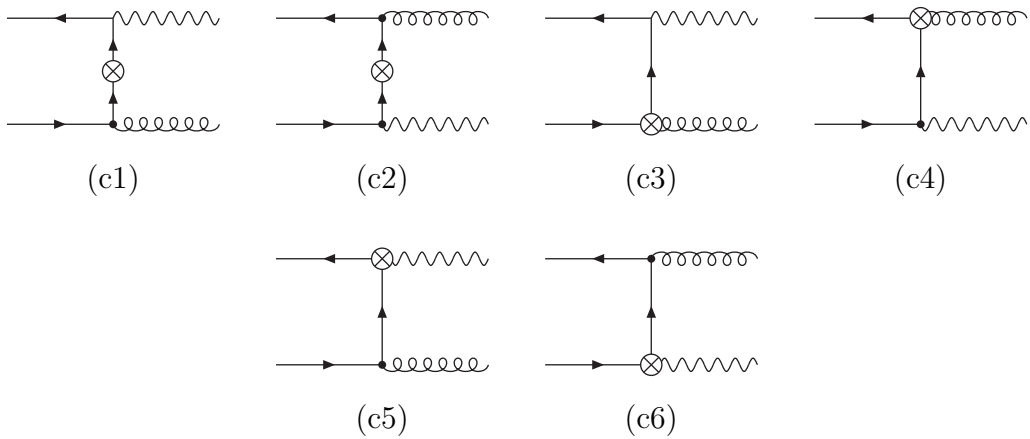

\begin{center}
\diagcI{(c1)}
\diagcII{(c2)}
\diagcIII{(c3)}
\diagcIV{(c4)}
\diagcV{(c5)}
\diagcVI{(c6)}
\end{center}
 \caption{Counterterm diagrams for the process $\bar q q \to Z g$.}
 \label{fig:ctdiags}
 \end{figure}}

We do not include electromagnetic corrections, 
\ie we restrict ourselves to the virtual weak contributions of
 $\ord(\alpha^2\alpha_\rS)$.
In the kinematical region where the $Z$-boson 
transverse momentum is non-vanishing
the emission of the gluon does not give rise to soft or collinear singularities
and the (renormalized) virtual weak corrections are finite.
All (real and virtual) quarks appearing in the diagrams of 
\reffi{fig:loopdiags} are treated as massless%
\footnote{
Quark-mass contributions are of $\ord(m_q^2/M_Z^2)$ and can thus be neglected
for $q\neq t$. 
The contributions of $\ord(m_t^2/M_Z^2)$ arising from those 
diagrams of \reffi{fig:loopdiags} that involve initial-state $b$ quarks and 
virtual $W$ bosons are large at the partonic level.
However, at the hadronic level, 
also these contributions can be neglected since 
the subprocesses initiated by $b$ quarks are suppressed by 
the very small $b$-quark density of protons.}
and diagrams involving couplings of quarks to 
Higgs bosons or would-be-Goldstone bosons are not considered.
Quark-mixing effects are also neglected.
The only quark-mass effects that we take into account are the 
$m_t$-and $m_b$-terms in the gauge-boson self-energies, 
which contribute to the counterterms.

Our calculation has been performed at the level of matrix elements
and provides full control over polarization effects.
However, at this level, the analytical expressions are too large to be published.
Explicit results will thus be presented only for the 
unpolarized squared matrix elements
\beqar\label{NLOunpol}
\overline{\sum}|\Mqq_{1}|^2 =
\overline{\sum}|\Mqq_{0}|^2 +
2\mathrm{Re}\,\left[\overline{\sum}\left(\Mqq_{0}\right)^*\, 
\de \Mqq_{1}\right]
+\ord(\alpha^3\alpha_\rS).
\eeqar

\subsection{Algebraic reduction}
\label{se:reduction}
The matrix element \refeq{oneloopsplitting} has the general form 
\beqar
\Mqq_1
&=&  \ri\,e\, g_\rS\,t^a\,  
\sum_{\lambda=\mathrm{R,L}} 
\bar{v}(p_{\qbar}) 
\M_1^{\la,\mu\nu} \omega_\la u(p_q)\, 
\varepsilon^*_\mu(p_Z) \varepsilon^*_\nu(p_g),
\eeqar
\nopagebreak
where the $\gamma^5$-terms are isolated in the chiral projectors 
$\omega_{\la}$ defined in \refeq{projectors}.
\pagebreak
Since we treat quarks as massless, 
$\M_1^{\la,\mu\nu}$ consists of terms involving an odd number of matrices 
$\gamma^\rho$ with $\rho=0,\dots,3$. 
The polarization dependence of the quark spinors 
and gauge-boson polarization vectors is implicitly understood.

In analogy to \refeq{oneloopsplitting}, \refeq{CTsplitting} we write
\beq
\M_1^{\la,\mu\nu}=\M_0^{\la,\mu\nu}+\de\M_1^{\la,\mu\nu}
,\qquad
\de\M_1^{\la,\mu\nu}=\de\M_{1,\mathrm{loops}}^{\la,\mu\nu}+\de\M_{1,\mathrm{CT}}^{\la,\mu\nu}
.
\eeq
The unpolarized squared matrix element \refeq{NLOunpol} is obtained from 
$\M_0^{\la,\mu\nu}$ and $\de\M_1^{\la,\mu\nu}$ using
\beqar\label{polsumid}
\overline{\sum}
\left( \Mqq_0\right)^* \de \Mqq_1
&=&
\pi^2 \alpha  \alpha_\rS (N_c^2-1)
\sum_{\lambda=\mathrm{R,L}} \mathrm{Tr}\left[
\ps_{q} \overline{\M}_0^{\la,\mu\nu}  \ps_{\qbar} \de \M_1^{\la,\mu'\nu'}\right] 
\nl&&{}\times
g_{\nu\nu'} \left( g_{\mu\mu'}-\frac{p_{Z\mu} p_{Z\mu'}}{M_Z^2} \right).
\eeqar
The Born contribution reads
\beqar\label{bornampli}
\M_0^{\la,\mu\nu}
&=& I^Z_{q_\lambda}
\smel^{\mu\nu}_0
,\qquad
\smel^{\mu\nu}_0=
\frac{\gamma^\mu(\ps_Z-\ps_\qbar)\gamma^\nu}{\that}+
\frac{\gamma^\nu(\ps_g-\ps_\qbar)\gamma^\mu}{\uhat},
\eeqar
and is simply proportional to the gauge group generator $I^Z$,
which determines the dependence of $\M_0$
on the weak mixing angle, 
the chirality ($\la=\rR,\rL$) 
and the weak isospin  of quarks ($u,d$).
Let us now consider  the combinations of 
gauge group generators that appear in the loop diagrams
of  \reffi{fig:loopdiags}.
The diagrams involving virtual $Z$ bosons are 
simply proportional to $(I^Z)^3$.
Instead,
the diagrams with virtual $W$ bosons (diagrams s1--b3 with $V=W^\pm$)
yield combinations of non-commuting
gauge-group generators  $I^Z,I^{W^\pm}$ and triple gauge-boson couplings
$\varepsilon^{W^\pm W^\mp Z}$, which can be expressed as%
\footnote{
The relation \refeq{couplings2} can be derived from 
Eqs.~(B.5) and (B.24) in \citere{Pozzorini:rs} whereas
\refeq{couplings3} follows from \refeq{couplings2}
combined with the commutation relation 
$[I^{W^\pm} , I^{Z}]=-(\ri/\sw) \varepsilon^{W^\pm W^\mp Z} I^{W^\pm}$.
}
\beqar
\label{couplings1}
&&I^{Z} I^{W^\pm} I^{W^\mp}
=
I^{W^\pm} I^{W^\mp}I^{Z}
\qquad\mbox{(diagrams  s1, s2, v1, v2)}
,\\
\label{couplings2}
&&\frac{\ri}{\sw} \varepsilon^{W^\pm W^\mp Z} I^{W^\pm} I^{W^\mp}
= 
\frac{\cw}{\sw^3} T^3
\qquad\mbox{(diagrams v5, v6, b3)}
,\\
\label{couplings3}
&&I^{W^\pm} I^{Z} I^{W^\mp} 
= 
I^{Z} I^{W^\pm} I^{W^\mp} 
-\frac{\cw}{\sw^3} T^3
\qquad\mbox{(diagrams v3, v4, b1, b2)}
,
\eeqar
where fermionic indices of the generators $I^V$
as well as summations over $W^\pm$ are implicitly understood.
The contribution of the loop diagrams of \reffi{fig:loopdiags} 
can thus be expressed as a linear combination of
$I^Z$ and $T^3$,
\beqar\label{ymsplitting}
\delta \M_{1,\mathrm{loops}}^{\la,\mu\nu}
&=& \frac{\alpha}{4\pi}\left[
I^Z_{q_\lambda}
  \sum_{V=\mathrm{Z,W^{\pm}}} \left(I^V I^{\bar{V}}\right)_{q_{\lambda}}
\delta \A_{1,\mathrm{A}}^{\mu\nu}(M^2_V)
+ \frac{\cw}{\sw^3} T^3_{q_\lambda}\, 
\delta \A_{1,\mathrm{N}}^{\mu\nu}(M^2_W)
\right]
,
\eeqar
with
\beq\label{couplfact1}
\left(I^Z I^{Z}\right)_{q_{\lambda}}
=
\left(I^Z _{q_{\lambda}}\right)^2
,\qquad
\sum_{V=W^{\pm}} \left(I^V I^{\bar{V}}\right)_{q_{\lambda}}
\equiv
\left(I^{W^\pm} I^{W^\mp}\right)_{q_{\lambda}}
=
\frac{\de_{\la\rL}}{2\sw^2}.
\eeq
The amplitudes $\delta \A_{1,\mathrm{A/N}}^{\mu\nu}(M^2_V)$
are denoted as the Abelian (A) and non-Abelian (N) contributions, 
since the latter originates from the non-commutativity of weak interactions,
whereas the former is present also in Abelian theories.

These amplitudes have been reduced algebraically using 
the Dirac equation, the identity
$p^\mu \varepsilon_\mu(p)=0$ for gauge-boson polarization vectors
and Dirac algebra.
Moreover, tensor loop integrals have been reduced to scalar ones 
by means of the Passarino-Veltman technique.
The result has been expressed in the form 
\beqar\label{algebraicred}
\delta \A_{1,\mathrm{A/N}}^{\mu\nu}(M^2_V)
=\sum_{i=1}^{10}  
\sum_{j=0}^{14}
\F_{\mathrm{A/N}}^{ij}(M_V^2)
\smel_{i}^{\mu\nu} 
\loops_{j}(M_V^2)
,
\eeqar
where the form factors $\F_{\mathrm{A/N}}^{ij}(M_V^2)$ 
are rational functions of Mandelstam invariants and masses.
The tensors 
\beqar
\smel^{\mu\nu}_{1} &=& \gamma^\mu(\ps_Z-\ps_{\qbar})\gamma^\nu
,\nl
\smel^{\mu\nu}_{2} &=&(\ps_Z-\ps_g)g^{\mu\nu}
,\nl
\smel^{\mu\nu}_{3} &=&  \gamma^\mu p_Z^\nu
,\nl
\smel^{\mu\nu}_{4} &=&-  \gamma^\nu p_g^\mu 
,\nl
\smel^{\mu\nu}_{5} &=&  \gamma^\mu  p_q^\nu
,\nl
\smel^{\mu\nu}_{6} &=&-   \gamma^\nu p_{\qbar}^\mu
,\nl
\smel^{\mu\nu}_{7} &=& (\ps_Z-\ps_g) p_g^\mu p_Z^\nu
,\nl
\smel^{\mu\nu}_{8} &=& (\ps_Z-\ps_g) p_{\qbar}^\mu p_q^\nu
,\nl
\smel^{\mu\nu}_{9} &=& (\ps_Z-\ps_g) p_g^\mu p_q^\nu
,\nl
\smel^{\mu\nu}_{10}&=&  (\ps_Z-\ps_g) p_{\qbar}^\mu p_Z^\nu,
\eeqar
correspond to the massless subset of the standard matrix elements 
of \citere{Denner:1991kt} and, apart from $\loops_{0}(M_V^2)=1$, 
the functions $\loops_{j}(M_V^2)$ represent the scalar one-loop integrals resulting from the reduction.%
\footnote{
For the scalar integrals $A_0,B_0,C_0$ and $D_0$
we adopt the notation of {\tt FeynCalc} \cite{Mertig:1990an}.
However, we choose their normalization according to \citere{Denner:1991kt}, 
\ie we include the factor $(2\pi\mu)^{4-D}$ which is omitted in the 
conventions of {\tt FeynCalc}.}
The one- and two-point functions,
\beqar\label{loopinta}
\loops_{ 1}(M_V^2) &=& A_0(M_V^2)
,\nl
\loops_{ 2}(M_V^2) &=& B_0(M_Z^2;0,0)
,\nl
\loops_{ 3}(M_V^2) &=& B_0(M_Z^2;M_V^2,M_V^2)
,\nl
\loops_{ 4}(M_V^2) &=& B_0(\shat;0,0)
,\nl
\loops_{ 5}(M_V^2) &=& B_0(\uhat;M_V^2,0)
,\nl
\loops_{ 6}(M_V^2) &=& B_0(\that;M_V^2,0)
,
\eeqar
are ultraviolet (UV) 
divergent. 
The resulting poles, in $D=4-2\varepsilon$ dimensions,
amount to
\beq\label{UVsing}
\left.\delta \A_{1,\mathrm{A}}^{\mu\nu}(M^2_V)\right|_{\mathrm{UV}}=
\frac{1}{\varepsilon}\smel_{0}^{\mu\nu} 
,\qquad
\left.\delta \A_{1,\mathrm{N}}^{\mu\nu}(M^2_V)\right|_{\mathrm{UV}}=
\frac{2}{\varepsilon}\smel_{0}^{\mu\nu}, 
\eeq
where $\smel_{0}^{\mu\nu}$ is the Born amplitude defined in \refeq{bornampli}.
These singularities are cancelled by the 
counterterms 
(see \refse{se:renormalisation}).
The remaining loop integrals are free from 
UV singularities.
The three-point functions 
\beqar\label{loopintb}
\loops_{ 7}(M_V^2) &=& C_0(\shat,0,0;0,0,M_V^2)
,\nl
\loops_{ 8}(M_V^2) &=& C_0(\uhat,M_Z^2,0;M_V^2,0,0)
,\nl
\loops_{ 9}(M_V^2) &=& C_0(\uhat,M_Z^2,0;0,M_V^2,M_V^2)
,\nl
\loops_{10}(M_V^2) &=& C_0(\that,M_Z^2,0;M_V^2,0,0)
,\nl
\loops_{11}(M_V^2) &=& C_0(\that,M_Z^2,0;0,M_V^2,M_V^2)
\eeqar
are finite for non-vanishing $Z$-boson transverse momentum.
In addition, we obtain the three- and four-point functions 
\beqar\label{singint}
&&
C_0(\uhat,0,0;M_V^2,0,0)
,\qquad
D_0(0,0,M_Z^2,0,\uhat,\shat;M_V^2,0,0,0)  
,\nl&&
C_0(\that,0,0;M_V^2,0,0)
,\qquad
D_0(0,0,M_Z^2,0,\that,\shat;M_V^2,0,0,0) 
,\nl&&
C_0(\shat,M_Z^2,0;0,0,0)
,\qquad
D_0(M_Z^2,0,0,0,\that,\uhat;M_V^2,M_V^2,0,0),
\eeqar
which result from the reduction of the box diagrams b1--b3.
As a consequence of vanishing quark masses, these latter integrals \refeq{singint} give rise to mass singularities of collinear nature.
Such singularities originate from the propagators of those massless (virtual) quarks that couple to the massless (real)
gluon,
\beq\label{collinearvertex}
\begin{picture}(40.,40.)(-20,-20)
\ArrowLine(-30,-20)(0,0)
\Gluon(0,0)(50,0){2.5}{6}
\ArrowLine(0,0)(-30,20)
\Vertex(-30,20){1.5}%
\Vertex(-30,-20){1.5}%
\Vertex(0,0){1.5}%
\Text(-12,-15)[l]{${k}$}
\Text(-12,15)[l]{${k-p_g}$}
\Text(30,-8)[t]{${p_g}$}
\Text(55,0)[c]{,}
\end{picture}
\eeq
specifically from the integration region with $k^\mu\to x p_g^\mu$,  
where the momenta of these quarks become collinear to the gluon momentum.
However, the box diagrams are finite since the quark-gluon vertex  \refeq{collinearvertex} yields
\beqar
(\ks-\ps_g) \varepss_g \ks 
\to 
x(x-1) \ps_g  \varepss_g \ps_g
= 
x(x-1) \left[ 2 \ps_g (p_g.\varepsilon_g)-\varepss_g p_g^2\right] =0,
\eeqar
in the collinear limit.
Indeed,
as a consequence of cancellations between the singularities 
from the $C_0$ and the $D_0$ functions in \refeq{singint}, 
our result is finite.

In order to control these cancellations at the analytical and numerical level,
we have regulated the singularities by means of an infinitesimal quark-mass parameter $\la$. 
Then, using \citere{Dittmaier:2003bc} we have expressed the singular parts of $D_0$ functions through singular $C_0$ functions and
checked that these cancel against the singular $C_0$ functions resulting from 
the reduction.
Finally we have written our result in terms 
of finite combinations of $D_0$ and $C_0$ functions,
\newcommand{\laa}{0}
\newcommand{\lab}{\la^2}
\newcommand{\lac}{0}
\beqar\label{loopintc}
\loops_{12}(M_V^2)
&=& 
D_0(0,0,M_Z^2,0,\uhat,\shat;M_V^2,\lac,\lab,\laa)  
-\frac{(\uhat-M_Z^2) C_0(\uhat,M_Z^2,0;M_V^2,\lab,0)}
{\shat\uhat+(\that+\uhat)M_V^2}
\nl&&{}
-\frac{
\uhat C_0(\uhat,0,0;M_V^2,\lab,\lac) 
+(\shat-M_Z^2) C_0(\shat,M_Z^2,0;\lac,\laa,\lab) }{\shat\uhat+(\that+\uhat)M_V^2},
\nl
\loops_{13}(M_V^2) 
&=&
D_0(0,0,M_Z^2,0,\that,\shat;M_V^2,\lac,\lab,\laa) 
-\frac{(\that-M_Z^2) C_0(\that,M_Z^2,0;M_V^2,\lab,0)}
{\shat\that+(\that+\uhat)M_V^2}
\nl&&{}
-\frac{\that C_0(\that,0,0;M_V^2,\lab,\lac) +
(\shat-M_Z^2) C_0(\shat,M_Z^2,0;\lac,\laa,\lab)}{\shat\that+(\that+\uhat)M_V^2} 
,\nl
\loops_{14}(M_V^2) 
&=&
D_0(M_Z^2,0,0,0,\that,\uhat;M_V^2,M_V^2,\lac,\lab)  
\nl&&{}
-\frac{\that C_0(\that,0,0;M_V^2,\lac,\lab)
+ \uhat C_0(\uhat,0,0;M_V^2,\lab,\lac)}{\that\uhat-(\that+\uhat)M_V^2} 
, 
\eeqar
which correspond to the finite parts of four-point integrals expressed as 
$D_0-D_0^{\mathrm{sing}}$, with the singular parts  $D_0^{\mathrm{sing}}$ in terms of $C_0$ functions.
It was checked that the functions $\loops_{12},\loops_{13},\loops_{14}$ are 
indeed finite and numerically stable for $\la/M_Z\ll 1$. 
The numerical results presented in \refse{se:numerics} have been obtained using  $\la/M_Z=10^{-6}$.

\subsection{Renormalization}
\label{se:renormalisation}
The renormalization of the process $\qbar q\to Z g$ is provided by the 
diagrams depicted in \reffi{fig:ctdiags}.
The counterterms (CTs)
 that are responsible for the contributions
of diagrams c1, c2, c3 and c4 read
\beq
\hspace{1.5cm}
\diagqqct
= 
\ri \ps \sum_{\lambda=\mathrm{R,L}}
\omega_{\lambda} \delta Z_{q_{\lambda}}
,\hspace{2cm}
\diagqqgct
= 
-\ri g_\rS  t^a \gamma^{\mu} \sum_{\la=\rR,\rL} 
\omega_{\la} \delta Z_{q_{\lambda}}
.
\eeq
Since there is no $\ord(\alpha)$-contribution to the renormalization 
of the strong coupling constant $g_\rS$,
these CTs depend only on the 
fermionic wave-function renormalization constants 
$\delta Z_{q_{\lambda}}$  [see \refeq{Qwfcts}].
Their combined contribution to the $\qbar q\to Z g$ process,
\ie the sum of diagrams  c1, c2, c3 and c4, vanishes.
The renormalization of the $\qbar q\to Z g$ process is thus 
provided by the diagrams c5, c6, which originate from 
the $Zq\qbar$ counterterm, 
\beqar\label{qqzct}
\diagqqzct
&=& \ri e \gamma^{\mu}
\sum_{\lambda=\mathrm{R,L}} \omega_{\lambda}  \left[I^{Z}_{q_{\lambda}} \, 
\left(\delta Z_{q_{\lambda}}+\delta C^{\mathrm{A}}_{q_\lambda}\right)
+ \frac{\cw}{\sw} T^3_{q_{\lambda}} \, \delta C^{\mathrm{N}}_{q_{\lambda}}
\right],
\eeqar
and yield
\beqar\label{Ctcontrib}
\de\M_{1,\mathrm{CT}}^{\la,\mu\nu}
&=& 
 \left[
I^{Z}_{q_{\lambda}} \, 
\left(\delta Z_{q_{\lambda}}+\delta C^{\mathrm{A}}_{q_\lambda}\right)
+ \frac{\cw}{\sw} T^3_{q_{\lambda}} \, \delta C^{\mathrm{N}}_{q_{\lambda}}
\right]
\smel_{0}^{\mu\nu},
\eeqar
with  $\smel_{0}^{\mu\nu}$ defined as in \refeq{bornampli}.
As indicated in  \refeq{qqzct}, the
$Zq\qbar$ CT can be expressed as a linear combination of 
$I^Z_{q_\la}$ and $T^3_{q_\la}$, analogous to \refeq{ymsplitting}.
The corresponding coefficients involve 
the wave-function renormalization constants of massless chiral quarks,%
\footnote{
$\Sigma^{q,{\la}}(p^2)$ are the chiral components of the 
one-particle irreducible two-point function of massless fermions,
\begin{displaymath}
\Gamma^{q\bar{q}}(p)
=
\ri \ps
\left[ 1+ \sum_{\la=\rR,\rL }\omega_\la  \Sigma^{q,{\la}}(p^2)\right]. 
\end{displaymath}
}
\beq\label{Qwfcts}
\delta Z_{q_{\lambda}} 
=
- \mathrm{Re}\left[\Sigma^{q,{\lambda}}(0)\right]
=
\frac{\alpha}{4\pi} \sum_{V=\mathrm{Z,W^\pm}}
\left( I^V I^{\bar{V}} \right)_{q_{\lambda}}
\left[\frac{3}{2} -\frac{A_0(M_V^2)}{M_V^2}\right],
\eeq
where only weak corrections ($V=\mathrm{Z,W^\pm}$) are included,
and the following combinations of renormalization constants
\beqar\label{deCCts}
\delta C^{\mathrm{A}}_{q_\lambda} 
&=&
\frac{1}{2} \left( \delta Z_{ZZ} +\frac{\cw}{\sw} \delta Z_{AZ} 
+ \frac{\delta e^2}{e^2} - \frac{1}{\sw^2} \frac{\delta \cw^2}{\cw^2}
\right)
,\nl
\delta C^{\mathrm{N}}_{q_{\lambda}} 
&=&
-\frac{1}{2\sw\cw} \delta Z_{AZ} + \frac{1}{\sw^2} \frac{\delta \cw^2}{\cw^2}.
\eeqar
The renormalization constants associated with the $Z$-boson wave function read
\beq\label{Zwfcts}
\delta Z_{ZZ} 
= 
-\mathrm{Re}
\left.
\frac{\partial
\Sigma^{{ZZ}}_{\rT}(p^2) 
}{\partial p^2} 
\right| 
_{p^2=M_Z^2}
,\qquad
\delta Z_{AZ} 
= 
-2 \mathrm{Re}
\left[ \frac{\Sigma^{\mathrm{AZ}}_{\mathrm{T}}(M_Z^2)}{M_Z^2}  \right]
\eeq
and have been evaluated using the self-energies of \citere{Denner:1991kt}.

For the renormalization of $e$ and $\cw$ we have 
considered two different schemes, the \msbar~and the on-shell scheme.
In the \msbar~scheme, the CTs for $e$ and $\cw$ read 
\newcommand{\Deltamsbar}{\bar \Delta_{\mathrm{UV}}}
\beqar\label{msbarcts}
\frac{\de \cw^2}{\cw^2}&\MSa&-\frac{\alpha}{4 \pi}\Deltamsbar
\left[\frac{19+22 \sw^2}{6 \cw^2}
+2(\rho-1)\right]
,\nl
\frac{\de e^2}{e^2}&\MSa&
\frac{\alpha}{4 \pi}
\Deltamsbar
\left[\frac{11}{3}
+\left(\frac{2}{\sw^2}-4\right)(\rho-1)\right]
,
\eeqar
where
\beq\label{msbarsubt}
\Deltamsbar = 1/\varepsilon -\gamma_{\mathrm{E}} +\log(4\pi)+\log\left(\frac{\mu_\rR^2}{M_Z^2}\right)
\eeq
in $D=4-2\varepsilon$ dimensions and $\rho=M_W^2/(\cw^2 M_Z^2)$.
In \refeq{msbarsubt} we have included a logarithmic term 
that renders the renormalized amplitude independent 
of the scale $\mu_\rR$ of dimensional regularization. 
This is equivalent to the choice $\mu_\rR=M_Z$ within the usual \msbar~scheme.
We also note that the counterterms \refeq{msbarcts} contain contributions 
proportional to $(\rho-1)$ which depend on the relation 
between $M_W$ and $M_Z$. 
In principle, in an $\ord(\alpha)$  \msbar~calculation the value of the $W$ mass,
which appears only in loop diagrams, should be derived from the on-shell $Z$-boson 
mass and the \msbar~mixing angle using the tree-level relation $M_W=\cw M_Z$.
In this case the $(\rho-1)$ contributions in \refeq{msbarcts}
would vanish.
However, it seems more natural to use the on-shell value 
of $M_W$ as an input parameter in our calculation.
This violates the tree-level mass relation and introduces
loop corrections that are implicitly contained 
in the numerical value of $M_W$. 
Their effect on our $\ord(\alpha)$ predictions is formally of $\ord{(\alpha^2)}$
since $M_W$ does not contribute at tree level.
This procedure is thus consistent at  $\ord(\alpha)$.
However, it affects also those  $1/\varepsilon$ poles 
that appear in $\de Z_{AZ}$ since they depend on $M_W$.
Such singularities give rise to terms proportional to $(\rho-1)/\varepsilon$
that are again, formally, of $\ord{(\alpha^2)}$ and are compensated by 
the $(\rho-1)$ terms that we have introduced in \refeq{msbarcts}.
This procedure removes all higher-order effects from the divergent part of the counterterms 
\refeq{deCCts}, which becomes independent of $(\rho-1)$,
and ensures the cancellation of the 
singularities \refeq{UVsing} originating from the loop diagrams
for arbitrary values of $\rho$.

In the on-shell (OS) scheme, the weak mixing angle is defined as
$\cw^2={M_W^2}/{M_Z^2}$ and the corresponding CT reads
\beq\label{weakmixct}
\frac{\delta \cw^2}{\cw^2} \OSa
\mathrm{Re}\left[ 
\frac{\Sigma^{WW}_{\rT}(M_W^2) }{M_W^2}
-\frac{\Sigma^{ZZ}_{\rT}(M_Z^2)}{M_Z^2} 
\right].
\eeq
As on-shell input parameter for the 
electromagnetic coupling constant we used 
{$\alpha=\alpha(M_Z^2)$}, defined as 
\beq
\alpha(M_Z^2)=\frac{\alpha(0)}{1-\Delta\alpha(M_Z^2)}
,\qquad
\Delta\alpha(M_Z^2)=\mathrm{Re}
\left[ \Pi^{AA}_{\mathrm{ferm}} (0)-\Pi^{AA}_{\mathrm{ferm}}(M_Z^2)\right]
,
\eeq
where $\alpha(0)=e(0)^2/(4\pi)$ is the fine-structure constant 
in the Thompson limit and 
$ \Pi^{AA}_{\mathrm{ferm}}$ represents the fermionic contribution
to the photonic vacuum polarization.
The CT associated with 
$e^2=4 \pi \alpha (M_Z^2)$ is given by
\beq\label{alphactos}
\frac{\delta e^2}{e^2}\OSa
\frac{\delta e(0)^2}{e(0)^2} -\Delta \alpha (M_Z^2)=
\mathrm{Re} \left[\Pi^{AA}_{\mathrm{fer}}(M_Z^2)\right]
+\Pi^{AA}_{\mathrm{bos}}(0)
- \frac{2\sw}{\cw} \frac{\Sigma^{AZ}_{\rT}(0)}{M_Z^2},
\eeq
where $\de e(0)$ is the CT in the Thompson limit,
and $\Pi^{AA}_{\mathrm{bos}}(0)$ represents the bosonic contribution 
to the photon propagator.

Independently  of the renormalization scheme,
the above CTs yield the ultraviolet singularities%
\footnote{
Note that $\delta C^{\mathrm{A}}_{q_\lambda}$ is free from 
ultraviolet  singularities,
\ie the pole associated with the 
Abelian coupling structure in \refeq{Ctcontrib} results exclusively 
from the fermionic wave-function renormalization constant $\delta Z_{q_\la}$ 
whereas the poles originating from the renormalization of 
the electroweak coupling constants $(\de e,\de\cw)$ and the 
$Z$-boson field renormalization constants $(\de Z_{ZZ},\de Z_{AZ})$
cancel in $\delta C^{\mathrm{A}}_{q_\lambda}$.
}
\beq
\left.\delta Z_{q_\lambda}\right|_{\mathrm{UV}} 
=-\frac{\alpha}{4\pi}\frac{1}{\varepsilon}
 \sum_{V=\mathrm{Z,W^\pm}}
\left( I^V I^{\bar{V}} \right)_{q_{\lambda}}
,\quad
\left.\delta C^{\mathrm{A}}_{q_\lambda}\right|_{\mathrm{UV}} 
=0
,\quad
\left.\delta C^{\mathrm{N}}_{q_\lambda}\right|_{\mathrm{UV}} 
=-\frac{\alpha}{2\pi\sw^2}\frac{1}{\varepsilon}
\eeq
that cancel the $1/\varepsilon$ poles 
\refeq{UVsing} originating from the loop diagrams.
The scheme dependence of the complete result is 
discussed in \refse{se:numerics}.

\subsection{Result}
\label{se:results}
Let us present the complete  $\ord(\alpha^2\alpha_\rS)$ result
for the unpolarized squared matrix element \refeq{NLOunpol}
for the $\qbar q\to Z g$ process.
Combining the contributions 
\refeq{bornampli}, \refeq{ymsplitting}, \refeq{Ctcontrib}, 
and using \refeq{polsumid} 
we obtain
\beqar\label{generalresult}
\lefteqn{
\overline{\sum}
|\M_{1}|^2 =
8 \pi^2 \alpha \alpha_\rS (N_\mathrm{c}^2-1)\,
}\quad
\nl&&{}\times      
\sum_{\lambda=\mathrm{R,L}}  
\Bigg\{  
\left(I^Z_{q_\la}\right)^2  \bigg[ 
H_0 \left( 1+2\delta C^{\mathrm{A}}_{q_{\lambda}} \right)  
+ \frac{\alpha}{2\pi}
\sum_{V=\mathrm{Z,W^{\pm}}} \left(I^V I^{\bar{V}}\right)_{q_{\lambda}}
\,H_1^{\mathrm{A}}(M_V^2)
\bigg]
\nl&&{}      
+ \frac{\cw}{\sw} T^3_{q_\la}I^Z_{q_\la} 
\bigg[
2 H_0\delta C^{\mathrm{N}}_{q_{\lambda}}
+ \frac{\alpha}{2\pi}
\frac{1}{\sw^2}H_1^{\mathrm{N}}(M_W^2)
\bigg]\Bigg\}
,
\eeqar
where the coupling factors are specified by
\refeq{couplfact0} and \refeq{couplfact1},
and
\beqar\label{h0res}
H_0&=&
\frac{1}{8}\mathrm{Tr}\left[ \,
\ps_{q} \, \overline{\smel}_0^{\mu\nu} \, \ps_{\qbar} \, \smel_0^{\mu'\nu'} \, 
\right] 
g_{\nu\nu'} \left( g_{\mu\mu'}-\frac{p_{Z\mu} p_{Z\mu'}}{M_Z^2} \right)
=
\frac{ \that^2+\uhat^2+2\shat M_Z^2}{\that\uhat}
\eeqar
represents the Born contribution.
The counterterms 
$\delta C^{\mathrm{A}/\mathrm{N}}_{q_{\lambda}}$
are defined in 
\refeq{deCCts}--\refeq{alphactos}. 
$H_1^{\mathrm{A/N}}(M_V^2)$
are the Abelian (A) and non-Abelian (N) contributions 
resulting from the loop diagrams of \reffi{fig:loopdiags}
and the fermionic wave-function renormalization constants \refeq{Qwfcts}.
These functions can be written as
\beqar
   H_1^{\mathrm{A/N}}(M_V^2) &=& 
\mathrm{Re}\left\{ \sum_{j=0}^{14} 
\left[
K_j^{\mathrm{A/N}}(M_V^2)
+K_{\mathrm{WF},j}^{\mathrm{A/N}}(M_V^2)
\right]
J_j(M_V^2)\right\},
\eeqar
where $J_0(M_V^2)=1$ and the remaining $J_j(M_V^2)$
are the loop integrals defined in 
\refeq{loopinta}, 
\refeq{loopintb},
 \refeq{loopintc}.
The fermionic wave-function renormalization constants 
yield [cf.~\refeq{Qwfcts}]
\beq\label{ferWFcont}
K_{\mathrm{WF},0}^{\mathrm{A}}(M_V^2)=\frac{3}{2}H_0,\qquad
K_{\mathrm{WF},1}^{\mathrm{A}}(M_V^2)=-\frac{1}{\MV^2}H_0,
\eeq
and $K_{\mathrm{WF},i}^{\mathrm{A/N}}(M_V^2)=0$ otherwise. 
The coefficients resulting from the loop integrals of \reffi{fig:loopdiags}
read
\beqar\label{implcoeff}
K_j^{\mathrm{A/N}}(M_V^2)&=&
\frac{1}{8}\sum_{i=1}^{10}  
\F^{ij}_{\mathrm{A/N}}(M_V^2)\,
\mathrm{Tr}\left[ \,
\ps_{q} \, \overline{\smel}_0^{\mu\nu} \, \ps_{\qbar} \, \smel_i^{\mu'\nu'} \, 
\right] 
g_{\nu\nu'} \left( g_{\mu\mu'}-\frac{p_{Z\mu} p_{Z\mu'}}{M_Z^2} \right),
\eeqar
where $\F^{ij}_{\mathrm{A/N}}(M_V^2)$ are the form factors of \refeq{algebraicred}.
The explicit expressions for these coefficients 
are presented in \refapp{app:coeff}.

In fact, as a consequence of Ward identities
(see \refse{se:checks}), the contributions of the
$p_{Z\mu} p_{Z\mu'}/M_Z^2$ terms in \refeq{h0res} and \refeq{implcoeff}
are equal zero.
We note that the contributions of right- and left-handed quarks
to the unpolarized squared matrix element \refeq{generalresult} 
can be easily recognized 
as the terms with $\la=\rR$ and $\la=\rL$, respectively.

\subsection{High-energy limit}
\label{se:helimit}
In this section we discuss the behaviour of the corrections \refeq{generalresult} to the $\qbar q\to Z g$ process in the high-$p_\rT$ region. 
More specifically, we derive approximate expressions for the 
functions $H_1^{\mathrm{A/N}}$ and the counterterms
$\delta C^{\mathrm{A/N}}_{q_{\lambda}}$
for the case where all energy scales are much higher than the weak-boson mass scale, \ie in the limit where $M_W^2/\shat\to 0$ and $\that/\shat$, $\uhat/\shat$ are constant.

In this approximation, 
which is applicable for transverse momenta of order of 100 GeV and beyond, 
the one-loop  weak corrections are strongly enhanced by logarithms of the form
$\log(\shat/M_W^2)$. The logarithmically enhanced part of the 
corrections, which constitutes the so-called 
next-to-leading logarithmic (NLL) approximation,
was already presented in \citere{Kuhn:2004em}.
This NLL part consists of double- and single-logarithmic
terms
that result  from the functions
$H_1^{\mathrm{A/N}}$ in \refeq{generalresult}
and read%
\footnote{
The correspondence between \refeq{generalresult}, \refeq{nllapprox}
and (5), (10) of \citere{Kuhn:2004em}
can be easily seen by means of the relation 
\begin{displaymath}
\sum_{V=\mathrm{Z,W^{\pm}}} \left(I^V I^{\bar{V}}\right)_{q_{\lambda}}
=C^{\mathrm{ew}}_{q_{\lambda}}-Q^2_q
\end{displaymath}
where $C^{\mathrm{ew}}_{q_{\lambda}}$ is the electroweak Casimir operator
and the $-Q^2_q$ term represents the subtraction of the electromagnetic contributions \cite{Kuhn:2004em}.
We also recall that the result of \citere{Kuhn:2004em}
is based on the equal-mass approximation, $M_Z=M_W$. 
}
\cite{Kuhn:2004em}
\beqar\label{nllapprox}
H_1^{\mathrm{N}}(M_W^2)&\NLLa&
-\left[
\log^2\left(\frac{|\that|}{M_W^2}\right)
+\log^2\left(\frac{|\uhat|}{M_W^2}\right)
-\log^2\left(\frac{|\shat|}{M_W^2}\right)
\right] H_0
,\nl
H_1^{\mathrm{A}}(M_V^2)&\NLLa&
-\left[\log^2\left(\frac{|\shat|}{M_W^2}\right)
-3\log\left(\frac{|\shat|}{M_W^2}\right)
\right] H_0,
\eeqar
whereas the counterterms
do not contribute to the NLL approximation, \ie 
\beq\label{nllapprox2}
\delta C^{\mathrm{A}}_{q_{\lambda}} \NLLa
\delta C^{\mathrm{N}}_{q_{\lambda}} \NLLa
0.
\eeq
The NLL corrections \refeq{nllapprox} are proportional to the 
Born contribution $H_0$ defined in \refeq{h0res}.
The above result is free from logarithms associated with the 
running of the electroweak coupling constants.
Also no single logarithms originating from
virtual collinear weak bosons that couple to the 
final-state $Z$ boson appear in \refeq{nllapprox}.
This is a general feature of gauge-boson 
production processes \cite{Denner:2001jv,Pozzorini:rs}.

In the following, we provide the complete 
asymptotic behaviour of the weak corrections, 
which includes, in addition to double and single logarithms,
also those contributions that are not logarithmically enhanced
in the high-energy limit.
This constitutes formally the next-to-next-to-leading logarithmic (NNLL) 
approximation and provides a precision of order 
$(M_W^2/\shat)\log^2(\shat/M_W^2)$.

Let us first consider the functions $H_1^{\mathrm{A/N}}$ in 
\refeq{generalresult}, 
which result from the loop diagrams of  \reffi{fig:loopdiags}
and the fermionic wave-function renormalization.  
The asymptotic behaviour of these functions was
derived using the general results of \citere{Roth:1996pd}
for the high-energy limit of one-loop integrals.
In addition, in order to simplify the dependence of the result
on the masses of the $Z$ and $W$ bosons,
we have performed an expansion in 
the Z-W mass difference 
using $\sw^2=1-M_W^2/M_Z^2$
as the expansion parameter and 
keeping only terms up to the first order in 
$\sw^2$.

The resulting NNLL expressions for  
$H_1^{\mathrm{N}}(M_W^2)$, 
$H_1^{\mathrm{A}}(M_W^2)$ and
$H_1^{\mathrm{A}}(M_Z^2)$ 
have the general form
\beq\label{nnllstracture}
H_1^{\mathrm{A/N}}(M_V^2)\NNLLa
\mathrm{Re }\,\left[
g_0^{\mathrm{A/N}}(M_V^2)\,
\frac{ \that^2+\uhat^2}{\that\uhat}
+g_1^{\mathrm{A/N}}(M_V^2)\,
\frac{ \that^2-\uhat^2}{\that\uhat}
+g_2^{\mathrm{A/N}}(M_V^2)
\right],
\eeq
\ie they involve the function
$(\that^2+\uhat^2)/\that\uhat$,
which corresponds to the high-energy limit of the 
Born contribution $H_0$ [see \refeq{h0res}],
and two other rational functions of the Mandelstam invariants,
which describe different angular dependences.
The functions $g_i^{\mathrm{A/N}}$ consist of logarithms 
of the kinematical variables and constants.
For the non-Abelian contribution $H_1^{\mathrm{N}}(M_W^2)$ we obtain
\beqar\label{heres1}
 g_0^{\mathrm{N}}(M_W^2)&=&
2\Deltamsbar
+\log^2\left(\frac{-\shat}{M_W^2}\right)
-\log^2\left(\frac{-\that}{M_W^2}\right)
-\log^2\left(\frac{-\uhat}{M_W^2}\right)
+\log^2\left(\frac{\that}{\uhat}\right)
\nl&&{}
-\frac{3}{2}\Biggl[
\log^2\left(\frac{\that}{\shat}\right)
+\log^2\left(\frac{\uhat}{\shat}\right)
\Biggr]
-\frac{20\pi^2}{9}
-\frac{2\pi}{\sqrt{3}}
+4 
,\nl
 g_1^{\mathrm{N}}(M_W^2)&=&
\frac{1}{2}\Biggl[
\log^2\left(\frac{\uhat}{\shat}\right)
-
\log^2\left(\frac{\that}{\shat}\right)
\Biggr]
,\nl
 g_2^{\mathrm{N}}(M_W^2)&=&
-2\Biggl[
\log^2\left(\frac{\that}{\shat}\right)
+\log^2\left(\frac{\uhat}{\shat}\right)
+\log\left(\frac{\that}{\shat}\right)
+\log\left(\frac{\uhat}{\shat}\right)
\Biggr]
-4{\pi^2}
,
\eeqar
where $\Deltamsbar$ is defined in \refeq{msbarsubt},
and  $H_1^{\mathrm{A}}(M_Z^2)$ is absent in \refeq{generalresult}.
For the Abelian contributions $H_1^{\mathrm{A}}(M_V^2)$ with 
$V=Z,W$ we obtain
\beqar\label{heres2}
 g_0^{\mathrm{A}}(M_V^2)&=&
-\log^2\left(\frac{-\shat}{M_V^2}\right)
+3\log\left(\frac{-\shat}{M_V^2}\right)
+\frac{3}{2}\Biggl[
\log^2\left(\frac{\that}{\shat}\right)
+\log^2\left(\frac{\uhat}{\shat}\right)
\nl&&{}
+\log\left(\frac{\that}{\shat}\right)
+\log\left(\frac{\uhat}{\shat}\right)
\Biggr]
+\frac{7\pi^2}{3}-\frac{5}{2}
,\nl
 g_1^{\mathrm{A}}(M_V^2)&=&
- g_1^{\mathrm{N}}(M_W^2)+
\frac{3}{2}\Biggl[
\log\left(\frac{\uhat}{\shat}\right)
-
\log\left(\frac{\that}{\shat}\right)
\Biggr]
,\nl
 g_2^{\mathrm{A}}(M_V^2)&=&
- g_2^{\mathrm{N}}(M_W^2).
\eeqar
The results \refeq{heres1}--\refeq{heres2}, for the 
$\qbar q\to Z g$ process, are valid for arbitrary values 
of the Mandelstam invariants. 
They can thus be easily translated to all other processes in \refeq{processes}
by means of appropriate permutations of $\shat,\that$ and $\uhat$,
which are specified by the crossing relations \refeq{crossing1} and \refeq{crossing2}.
We note that the arguments of the logarithms 
in \refeq{heres1}--\refeq{heres2} are in general not positive. 
Logarithms with negative arguments 
are defined through the 
usual $\ri\varepsilon$ prescription, 
$\rhat\to \rhat +\ri\varepsilon$
for  $\rhat=\shat,\that,\uhat$.
In practice we need only the real parts of the logarithms,
\beqar\label{logcont}
\mathrm{Re }\, \log(x)&=& \log(|x|)
,\nl
\mathrm{Re }\, \log^2(x)&=& \log^2(|x|)-\pi^2\theta(-x).
\eeqar
The $\pi^2$-terms originating from the analytic continuation of the double 
logarithms depend on the signs of their arguments
and these latter are determined by the 
signs of the Mandelstam invariants.
For the  $\qbar q\to Z g$ process, one has
$\shat>0,\that<0$ and $\uhat<0$.
However, these signs change when the 
Mandelstam invariants are permuted 
in order to translate the results  \refeq{heres1}--\refeq{heres2}
for the $\qbar q\to Z g$ process
to the other partonic processes in \refeq{processes}.
The $\pi^2$ terms resulting from \refeq{logcont}
are thus process dependent.
Simple explicit expression
that apply to all processes in \refeq{processes}
can be obtained using the relations 
\beqar\label{signrelatin}
\theta(\shat)+\theta(\that)+\theta(\uhat)
&=&1
,\qquad
\theta(\shat)\theta(\that)=\theta(\that)\theta(\uhat)=\theta(\uhat)\theta(\shat)=0
,
\eeqar
which are valid for $\shat>0$, $\that<0$ and $\uhat<0$ as well as for any 
permutation of the Mandelstam invariants (crossing transformations).
Combining \refeq{heres1}--\refeq{signrelatin} we obtain
\beqar\label{heres1b}
\mathrm{Re }\, g_0^{\mathrm{N}}(M_W^2)&=&
2\Deltamsbar
+\log^2\left(\frac{|\shat|}{M_W^2}\right)
-\log^2\left(\frac{|\that|}{M_W^2}\right)
-\log^2\left(\frac{|\uhat|}{M_W^2}\right)
+\log^2\left(\frac{|\that|}{|\uhat|}\right)
\nl&&{}
-\frac{3}{2}\Biggl[
\log^2\left(\frac{|\that|}{|\shat|}\right)
+\log^2\left(\frac{|\uhat|}{|\shat|}\right)
\Biggr]
-\frac{\pi^2}{18}\left[4+9\theta(-\shat)\right]
-\frac{2\pi}{\sqrt{3}}
+4 
,\nl
\mathrm{Re }\, g_1^{\mathrm{N}}(M_W^2)&=&
\frac{1}{2}\Biggl[
\log^2\left(\frac{|\uhat|}{|\shat|}\right)
-
\log^2\left(\frac{|\that|}{|\shat|}\right)
\Biggr]
+\frac{\pi^2}{2}\left[\theta(\that)-\theta(\uhat)
\right]
,\nl
\mathrm{Re }\, g_2^{\mathrm{N}}(M_W^2)&=&
-2\Biggl[
\log^2\left(\frac{|\that|}{|\shat|}\right)
+\log^2\left(\frac{|\uhat|}{|\shat|}\right)
+\log\left(\frac{|\that|}{|\shat|}\right)
+\log\left(\frac{|\uhat|}{|\shat|}\right)
\Biggr]
\nl&&
{}-2{\pi^2}\theta(-\shat),
\eeqar
and
\beqar\label{heres2b}
\mathrm{Re }\, g_0^{\mathrm{A}}(M_V^2)&=&
-\log^2\left(\frac{|\shat|}{M_V^2}\right)
+3\log\left(\frac{|\shat|}{M_V^2}\right)
+\frac{3}{2}\Biggl[
\log^2\left(\frac{|\that|}{|\shat|}\right)
+\log^2\left(\frac{|\uhat|}{|\shat|}\right)
\nl&&{}
+\log\left(\frac{|\that|}{|\shat|}\right)
+\log\left(\frac{|\uhat|}{|\shat|}\right)
\Biggr]
+\frac{\pi^2}{6}\left[2+3\theta(-\shat)\right]
-\frac{5}{2}
,\nl
\mathrm{Re }\, g_1^{\mathrm{A}}(M_V^2)&=&
- \mathrm{Re }\,g_1^{\mathrm{N}}(M_W^2)+
\frac{3}{2}\Biggl[
\log\left(\frac{|\uhat|}{|\shat|}\right)
-
\log\left(\frac{|\that|}{|\shat|}\right)
\Biggr]
,\nl
\mathrm{Re }\, g_2^{\mathrm{A}}(M_V^2)&=&
- \mathrm{Re }\,g_2^{\mathrm{N}}(M_W^2).
\eeqar
For the $\qbar q\to Z g$  
and $q \qbar\to Z g$ processes, 
with $\shat>0,\that<0$ and $\uhat<0$,
all $\theta$ functions in  \refeq{heres1b}--\refeq{heres2b} vanish. 
The  $\theta$-terms become relevant when the 
above results 
are translated to the processes that involve gluons 
in the initial state. 

Let us compare the double and single logarithms of $M_{V}$
that appear in  \refeq{heres1b}--\refeq{heres2b}
with the NLL result of \citere{Kuhn:2004em},
which is given in \refeq{nllapprox}.
In the  $M_V^2/\shat\to 0$ limit the two results agree 
up to contributions resulting from the difference between 
$g_0^{\mathrm{A}}(M_W^2)$ and $g_0^{\mathrm{A}}(M_Z^2)$.
Since the calculation of  \citere{Kuhn:2004em} 
was based on the equal-mass approximation, $M_W=M_Z$, we conclude 
that the two results are consistent.
\newcommand{\expnot}[1]{\times 10^{#1}}

Let us now consider the contribution of the 
counterterms (CTs)
 $\delta C^{\mathrm{A}/\mathrm{N}}_{q_{\lambda}}$ in
\refeq{generalresult}.
These CTs [see \refeq{deCCts}--\refeq{alphactos}]
consist of on-shell gauge-boson 
self-energies and  are independent 
of the high-energy scales $\shat$, $\that$ and $\uhat$. 
They are thus free from large logarithms.
Since the exact analytical expressions for the CTs are
relatively complicated as compared to the compact NNLL
contributions \refeq{heres1b}--\refeq{heres2b},
we present simple approximate expressions for
$\delta C^{\mathrm{A}/\mathrm{N}}_{q_{\lambda}}$.
Restricting ourselves to the case of the  \msbar~scheme,
we use:
(i) A heavy-top expansion including terms of order 
$\log(m_t)$ and $(1/m_t)^0$;
(ii) A first-order expansion in the Z-W mass difference, using 
$1-M_W^2/M_Z^2=\sw^2$ as expansion parameter;
(iii) The light-Higgs approximation $M_H=M_Z$.
The resulting expressions for the renormalization constants 
associated with the $Z$-boson wave function
read
\beqar\label{approxWFRRS}
\delta \tilde Z_{AZ}&=&
\frac{\alpha}{4\pi}
\frac{1}{\sw\cw}
\Biggl[
-\left(\frac{7+34\sw^2}{3}
+4\cw^2(\rho-1)
\right)\Deltamsbar
-\frac{4}{9}(3-8\sw^2)\log\left(\frac{m_t^2}{M_Z^2}\right)
\nl&&{}
+(17-36\sw^2)\frac{\pi}{\sqrt{3}}
-\frac{8}{27}(78-89\sw^2)
\Biggr]
,
\nl
\delta \tilde Z_{ZZ}&=&
\frac{\alpha}{4\pi}
\frac{1}{\sw^2}
\Biggl[
-\frac{5-10\sw^2+46\sw^4}{6\cw^2}
\Deltamsbar
+\frac{3-5\sw^2}{6}\log\left(\frac{m_t^2}{M_Z^2}\right)
\nl&&{}
+\frac{11-37\sw^2}{6}\frac{\pi}{\sqrt{3}}
-\frac{113-573\sw^2}{36}
\Biggr]
.
\eeqar
And for the CTs \refeq{deCCts}
within  the \msbar~scheme we obtain
\beqar\label{approxdeCCts}
\delta \tilde C^{\mathrm{A}}_{q_\lambda} 
&\MSa&
\frac{\alpha}{4\pi}
\frac{1}{\sw^2}
\Biggl[
-\frac{15-49\sw^2}{36}\log\left(\frac{m_t^2}{M_Z^2}\right)
\nl&&{}
+\frac{113-253\sw^2}{12}\frac{\pi}{\sqrt{3}}
-\frac{105}{8}
+\frac{4567}{216}\sw^2
\Biggr],\nl
\delta \tilde C^{\mathrm{N}}_{q_{\lambda}} 
&\MSa&
-\frac{\alpha}{4\pi}
\Biggl\{
\frac{2}{\sw^2}
\Deltamsbar
+\frac{1}{2\sw^2\cw^2}
\Biggl[
-\frac{4}{9}(3-8\sw^2)\log\left(\frac{m_t^2}{M_Z^2}\right)
\nl&&{}
+(17-36\sw^2)\frac{\pi}{\sqrt{3}}
-\frac{8}{27}(78-89\sw^2)
\Biggr]\Biggr\}
.
\eeqar
As one can easily infer from the numerical values%
\footnote{
These numerical values are obtained using the \msbar~input parameters
given in  \refse{se:numerics} and $\Deltamsbar=0$.
}
$\delta Z_{AZ} =-2.316\expnot{-3}$,
$\delta   Z_{ZZ} =3.943\expnot{-3}$,
$\delta\tilde  Z_{AZ} =-2.888\expnot{-3}$,
$\delta\tilde    Z_{ZZ}=4.533\expnot{-3}$,
the contribution of the CTs to the cross section 
is of order of a few permille 
whereas the error associated with the approximations
\refeq{approxWFRRS},\refeq{approxdeCCts}
is below the one permille level.

\subsection{Checks}
\label{se:checks}
In order to control the correctness of our results
we performed various consistency checks.
We have verified that the one-loop corrections \refeq{algebraicred}
satisfy the Ward Identities
\beqar
\varepsilon^*_\mu(p_Z)\,
p_{g\nu}\,
\bar{v}(p_{\qbar}) 
\left[\delta \A_{1,\mathrm{A/N}}^{\mu\nu}(M^2_V)
\omega_\la \right]
u(p_q)
&=&0
,\nl
p_{Z\mu}\,
\varepsilon^*_\nu(p_g) \,
\bar{v}(p_{\qbar}) 
\left[\delta \A_{1,\mathrm{A/N}}^{\mu\nu}(M^2_V)
\omega_\la \right]
u(p_q)
&=&0
,\nl
p_{Z\mu} \,p_{g\nu}
\,
\bar{v}(p_{\qbar}) 
\left[\delta \A_{1,\mathrm{A/N}}^{\mu\nu}(M^2_V)
\omega_\la \right]
u(p_q)
&=&0
.
\eeqar
Similar Ward identities hold for the lowest-order amplitude.
The cancellation of the ultraviolet  divergences 
and the compensation of the fictitious collinear singularities 
discussed in \refse{se:reduction}
have been verified analytically  and numerically.
The high-energy limit of our result has been worked out analytically 
and we have shown that the leading- and next-to-leading logarithmic 
contributions agree with \citere{Kuhn:2004em}.

Moreover, the calculation has been performed in two completely independent ways, 
based on different computer-algebra and numerical tools. 
On one hand we have written algorithms for the algebraic reduction in {\tt Mathematica} \cite{mathematica} and used a set of routines provided by A.~Denner for the numerical evaluation of loop integrals.
On the other hand we have performed the reduction by means of {\tt FeynCalc} \cite{Mertig:1990an} and used {\tt FF} \cite{vanOldenborgh:1990yc} to 
evaluate loop integrals.
The results have been implemented in different {\tt Fortran} codes 
and comparing them we find very good agreement at numerical level.

\section{Predictions for the hadronic Z production at high transverse
  momentum}
\label{se:numerics}

In the following we discuss numerical predictions for the 
quantities calculated above. The lowest order (LO) and 
the next-to-leading-order (NLO) predictions result
from \refeq{generalamplitude} and \refeq{generalresult}--\refeq{ferWFcont}, \refeq{nloabcoeff}--\refeq{nlonabcoeff}, respectively.
The next-to-leading-logarithmic 
(NLL) approximation, at one loop, 
corresponds%
\footnote{For details concerning the treatment of
angular-dependent logarithms 
at the NLL level we refer to \citere{Kuhn:2004em}.
}
to \refeq{nllapprox}--\refeq{nllapprox2}.
The next-to-next-to-leading-logarithmic (NNLL) 
predictions, also at one loop, are based on 
\refeq{nnllstracture}--\refeq{approxdeCCts}. 
All results are obtained using LO MRST2001 parton
distribution functions (PDFs) \cite{Martin:2002dr}.
We choose $\pT^2$ as the factorization scale
and, similarly as the scale at which the running strong coupling constant
is evaluated. We also adopt the value 
$\al_\rS(M_Z^2)=0.13$ and use the one-loop running expression for 
$\al_\rS(\mu^2)$, in accordance with the LO 
PDF extraction method
of the MRST collaboration. 
We use the 
following values of parameters~\cite{Eidelman:2004wy}: $M_Z=91.19 \GeV$,
\mbox{$M_W=80.39 \GeV$}, $m_t=176.9\GeV$, $m_b=4.3 \GeV$, $M_H=120 \GeV$. 
\pagebreak
For the calculation in the~\msbar~scheme we use
$\al=1/128.1$, $\sw^2=0.2314$, whereas for the
on-shell  scheme $\al=1/128.9$, $\sw^2=1-\cw^2=$
\linebreak
$1-M_W^2/M_Z^2$ are taken. 
\nopagebreak
The NLO predictions use the form of the 
counterterms
given by \refeq{deCCts}--\refeq{alphactos}, 
whereas the NNLL results are based on the 
approximate expressions~\refeq{approxWFRRS}--\refeq{approxdeCCts}
for the counterterms.
Unless otherwise noted, the 
results are obtained using the \msbar~scheme.

We begin by investigating NLO, NLL and NNLL relative corrections to
the partonic (unpolarized) differential cross section
${\rd \hat{\si}^{i j}}/{\rd \cos\theta}$ 
[see \refeq{partoniccs0}]. 
To this end we define 
\beq
\hat{{\cal R}}^{ij}_{\NLO/\LO}= 
{\rd \hat \si^{ij}_{\NLO}/\rd \cos\theta \over
 \rd \hat \si^{ij}_{\LO} /\rd \cos\theta} 
-1 
\eeq
and similarly ${\hat{\cal R}}^{ij}_{\NLL/\NLO}$ and 
$\hat{\cal R}^{ij}_{\NNLL/\NLO}$.   
\begin{figure}[]
\vspace*{2mm}
  \begin{center}
\epsfig{file=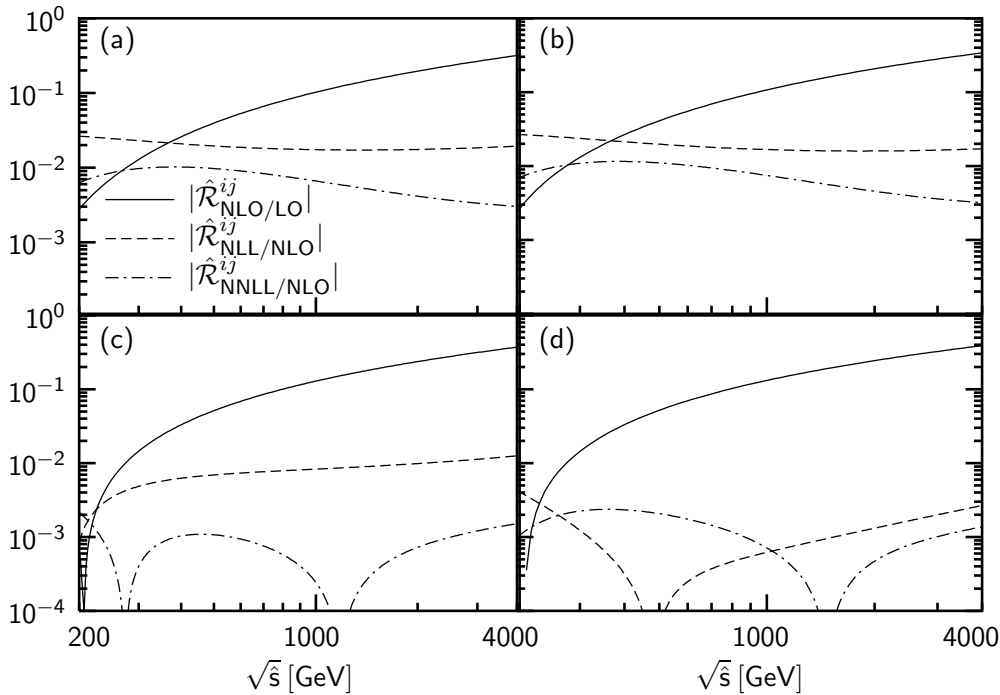, angle=0, width=13cm}
\end{center}
\vspace*{-2mm}
\caption{
Relative one-loop corrections to the partonic differential cross
sections $\rd \hat\si^{ij} / \rd \cos \theta$ at $\cos \theta =0$ 
for (a) $\bar u u$ channel,
(b) $ \bar d d$ channel, (c) $g u$ channel, (d) $ g d$ channel. The
solid, dashed and dot-dashed lines denote the 
modulus of the $\hat {\cal R}$ ratios, 
as defined in the text, for the full NLO cross section, the NLL
approximation and the NNLL approximation of the one-loop cross section, respectively. 
}
\label{fig:part}
\end{figure}
These ratios, calculated at  $\cos \theta =0$, are displayed as a
function of $\sqrt \shat$ in \reffi{fig:part}. We consider four
processes: $\bar u u \rar Zg$
(\reffi{fig:part}a), $\bar d d \rar Z g$ (\reffi{fig:part}b), $g
u \rar Z u$ (\reffi{fig:part}c), $g d \rar Z d$
(\reffi{fig:part}d). 
The size of the full weak NLO correction grows with 
the energy and reaches 30\% for all channels at $\sqrt{\shat}=4\TeV$.
The sign of this correction is negative.
From \reffi{fig:part}
we conclude that
the NLL terms provide a fairly good approximation to
the full NLO result  for $\sqrt{\shat}\ge 200\GeV$, 
with the remaining terms responsible for less
than 3\% of the cross section in the $u \bar u$ and $ d \bar d$
channels, and less than 1\% in the $g u$ and $g d$ channels. 
The quality 
of the NNLL approximation is very good
in all channels, 
better or comparable to 1\%
in the full region under consideration.
The absolute LO and NLO cross section 
and the $\hat{\cal R}^{ij}$ ratios 
for specific values of $\cos \theta$, $\sqrt \shat$ and different
collision channels are listed in Table~\ref{tab:part}. 
\begin{table}[]
\begin{tabular}{|c|c|c|l|l|r|r|r|}
\hline 
\multicolumn{3}{|c|}{} 
&\multicolumn{2}{|c|}{$\rd\hat\si^{ij} / \rd \cos \theta$ [pb] } 
&
\multicolumn{3}{|c|}{ $\hat{{\cal R}}_{\scriptscriptstyle{\mathrm{A}/\mathrm{B}}}^{ij}\times 10^3$} \\ \hline
$ij$
&$\frac{\sqrt{\hat{s}}}{\GeV}$ 
& $\cos \theta$ 
& \multicolumn{1}{|c|}{$\scriptstyle{\LO}$ }
& \multicolumn{1}{|c|}{$\scriptstyle{\NLO}$ }
& \multicolumn{1}{|c|}{ $\scriptscriptstyle{\NLO/\LO}$}
& \multicolumn{1}{|c|}{$\scriptscriptstyle{\NLL/\NLO}$}
& \multicolumn{1}{|c|}{$\scriptscriptstyle{\NNLL/\NLO}$}
\\ \hline
\hline
$\bar u u$
&   500 &  0.0 & $ 1.7012 \times 10^{  0}$ & $ 1.6341 \times 10^{  0}$ & $-39.431    $ & $19.501     $ & $9.7186     $\\\      
&  1000 &  0.0 & $ 3.5934 \times 10^{- 1}$ & $ 3.2261 \times 10^{- 1}$ & $-102.22    $ & $17.397     $ & $6.6197     $\\\      
&  2000 &  0.0 & $ 8.1081 \times 10^{- 2}$ & $ 6.5185 \times 10^{- 2}$ & $-196.05    $ & $17.518     $ & $4.1422     $\\\hline 
$\bar u u$
&   500 &  0.5 & $ 2.8203 \times 10^{  0}$ & $ 2.7210 \times 10^{  0}$ & $-35.224    $ & $25.906     $ & $7.2284     $\\\      
&  1000 &  0.5 & $ 6.0587 \times 10^{- 1}$ & $ 5.4862 \times 10^{- 1}$ & $-94.497    $ & $24.664     $ & $4.9420     $\\\      
&  2000 &  0.5 & $ 1.3718 \times 10^{- 1}$ & $ 1.1185 \times 10^{- 1}$ & $-184.66    $ & $25.881     $ & $3.2372     $\\\hline 
\hline 
$\bar d d$
&   500 &  0.0 & $ 2.1930 \times 10^{  0}$ & $ 2.1031 \times 10^{  0}$ & $-40.978    $ & $19.759     $ & $11.038     $\\\      
&  1000 &  0.0 & $ 4.6322 \times 10^{- 1}$ & $ 4.1349 \times 10^{- 1}$ & $-107.36    $ & $16.723     $ & $7.4723     $\\\      
&  2000 &  0.0 & $ 1.0452 \times 10^{- 1}$ & $ 8.2852 \times 10^{- 2}$ & $-207.31    $ & $16.002     $ & $4.5613     $\\\hline 
$\bar d d$
&   500 &  0.5 & $ 3.6357 \times 10^{  0}$ & $ 3.5013 \times 10^{  0}$ & $-36.970    $ & $25.829     $ & $8.3255     $\\\      
&  1000 &  0.5 & $ 7.8103 \times 10^{- 1}$ & $ 7.0278 \times 10^{- 1}$ & $-100.18    $ & $23.838     $ & $5.6448     $\\\      
&  2000 &  0.5 & $ 1.7684 \times 10^{- 1}$ & $ 1.4205 \times 10^{- 1}$ & $-196.74    $ & $24.376     $ & $3.5721     $\\\hline 
\hline 
$g u$
&   500 &  0.0 & $ 6.9404 \times 10^{- 1}$ & $ 6.5844 \times 10^{- 1}$ & $-51.284    $ & $-6.9286    $ & $-1.0724    $\\\      
&  1000 &  0.0 & $ 1.6267 \times 10^{- 1}$ & $ 1.4171 \times 10^{- 1}$ & $-128.83    $ & $-8.2907    $ & $-0.2447    $\\\      
&  2000 &  0.0 & $ 3.7676 \times 10^{- 2}$ & $ 2.8750 \times 10^{- 2}$ & $-236.92    $ & $-9.8681    $ & $0.7254     $\\\hline 
$g u$
&   500 &  0.5 & $ 5.9307 \times 10^{- 1}$ & $ 5.5148 \times 10^{- 1}$ & $-70.127    $ & $-13.649    $ & $0.9367     $\\\      
&  1000 &  0.5 & $ 1.3822 \times 10^{- 1}$ & $ 1.1654 \times 10^{- 1}$ & $-156.82    $ & $-16.198    $ & $1.0469     $\\\      
&  2000 &  0.5 & $ 3.1936 \times 10^{- 2}$ & $ 2.3182 \times 10^{- 2}$ & $-274.09    $ & $-19.650    $ & $1.4449     $\\\hline 
\hline 
$g d$
&   500 &  0.0 & $ 8.9468 \times 10^{- 1}$ & $ 8.4822 \times 10^{- 1}$ & $-51.931    $ & $-0.0567    $ & $-2.0723    $\\\      
&  1000 &  0.0 & $ 2.0969 \times 10^{- 1}$ & $ 1.8206 \times 10^{- 1}$ & $-131.76    $ & $-0.6210    $ & $-0.7120    $\\\      
&  2000 &  0.0 & $ 4.8569 \times 10^{- 2}$ & $ 3.6717 \times 10^{- 2}$ & $-244.01    $ & $-1.3163    $ & $0.4704     $\\\hline 
$g d$
&   500 &  0.5 & $ 7.6453 \times 10^{- 1}$ & $ 7.1421 \times 10^{- 1}$ & $-65.819    $ & $-5.1329    $ & $-0.1256    $\\\      
&  1000 &  0.5 & $ 1.7818 \times 10^{- 1}$ & $ 1.5103 \times 10^{- 1}$ & $-152.38    $ & $-6.6209    $ & $0.5373     $\\\      
&  2000 &  0.5 & $ 4.1168 \times 10^{- 2}$ & $ 2.9994 \times 10^{- 2}$ & $-271.41    $ & $-8.7165    $ & $1.1609     $\\\hline 
\end{tabular}
\caption{
Absolute value of the LO and NLO  partonic 
differential cross
section $\rd\hat\si^{ij} / \rd \cos \theta$ 
and the ratios  
$\hat{{\cal R}}^{ij}_{\NLO/\LO}$, $\hat{{\cal R}}^{ij}_{\NLL/\NLO}$ and 
$\hat{{\cal R}}^{ij}_{\NNLL/\NLO}$ in permille,
for the  partonic processes 
$\bar uu\to Z g$, 
$\bar dd\to Z g$, 
$gu\to Z u$ and
$gd\to Z d$.
The running strong coupling $\alpha_\rS(\mu^2)$
is taken at the scale $\mu^2=p_\rT^2=
(1-\cos^2\theta)(1-M_Z^2/\shat)^2\shat/4$.
}
\label{tab:part}
\end{table}

\begin{figure}[]
\vspace*{2mm}
  \begin{center}
\epsfig{file=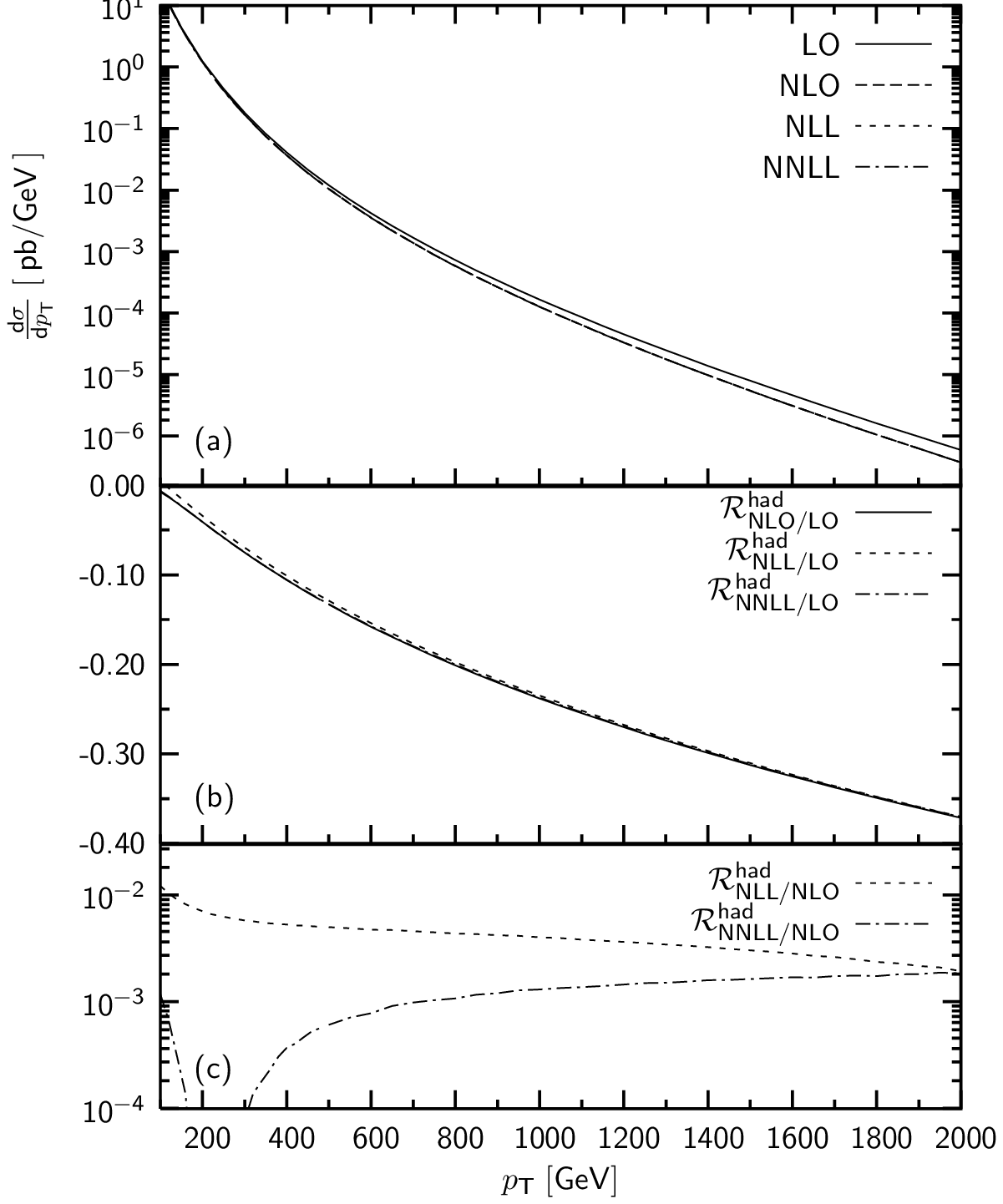, angle=0, width=11.5cm}
\end{center}
\vspace*{-2mm}
\caption{Transverse momentum distribution for $pp\rar Z j$ at
  $\sqrt{s}=14 \TeV$.
(a) LO (solid), NLO (dashed),  NLL (dotted) and 
NNLL (dot-dashed) predictions. 
(b) Relative NLO (solid), NLL (dotted) and NNLL (dot-dashed)
weak correction \wrt the LO distribution.
(c) NLL (dotted) and NNLL (dot-dashed) approximations compared to the 
full NLO result.
}
\label{fig:lhc}
\end{figure}

The  transverse momentum distribution $ \rd \si /
\rd\pT$ at the LHC is shown in \reffi{fig:lhc}. 
We display separately the
absolute values of the LO, NLO, NLL
and NNLL differential cross sections (\reffi{fig:lhc}a) and the relative
correction \wrt the LO result for the NLO, NLL and NNLL
distributions (\reffi{fig:lhc}b). The relative correction  \wrt
the LO is now defined as 
\beq
{{\cal R}}^{\rm had}_{\NLO/\LO}= 
{\rd  \si_{\NLO}/\rd \pT \over
 \rd  \si_{\LO} /\rd \pT} 
-1  
\eeq
for the NLO case, and similarly for the NLL and the NNLL cross sections. 
The quality of the NLL and NNLL high-energy  approximations 
is shown in more detail in \reffi{fig:lhc}c. The
importance of the NLO corrections increases significantly with $\pT$.
The NLO correction results in a negative contribution ranging from $-13\%$ at 
\mbox{$\pT=500 \GeV$} up to $-37\%$ at $\pT=2 \TeV$ of the LO cross section.
We observe that the NLL approximation works very well.
It differs from the full NLO prediction by about 
 1\% at low $\pT$ and by 0.2\% at $\pT\sim 2 \TeV$, cf. Fig~\ref{fig:lhc}c. 
The quality of the NNLL approximation 
is at the 
permille level (or better) in the entire $\pT$ range.

\begin{figure}[]
\vspace*{2mm}
  \begin{center}
\epsfig{file=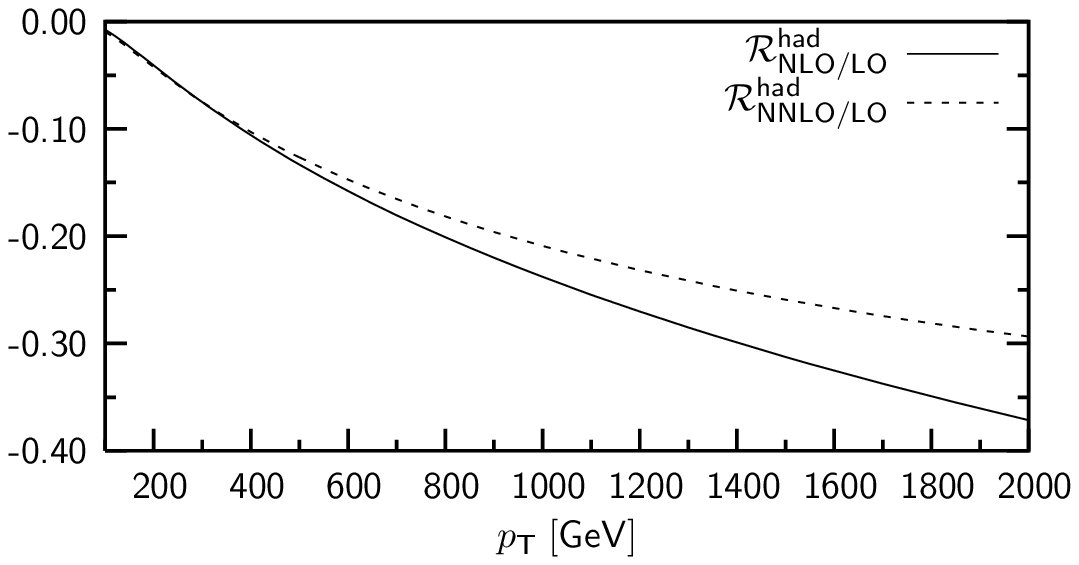, angle=0, width=11.5cm}
\end{center}
\vspace*{-2mm}
\caption{
Relative NLO (solid) and NNLO (dotted) corrections 
to the $p_\rT$ distribution for $pp\rar Z j$ at $\sqrt{s}=14 \TeV$.
}
\label{fig:lhcpt12}
\end{figure}

In view of the large one-loop effects, we include the dominant two-loop terms
 \cite{Kuhn:2004em}.
In \reffi{fig:lhcpt12} 
we show the relative size of the corrections 
in the NLO approximation (${{\cal R}}^{\rm had}_{\NLO/\LO}$, solid line), 
and in the approximation which includes the 
next-to-leading logarithmic two-loop terms
 (${{\cal R}}^{\rm had}_{\NNLO/\LO}$, dotted line).
These additional two-loop terms are positive and amount to 8\% for $p_\rT=2\TeV$.

\begin{figure}[]
\vspace*{2mm}
  \begin{center}
\epsfig{file=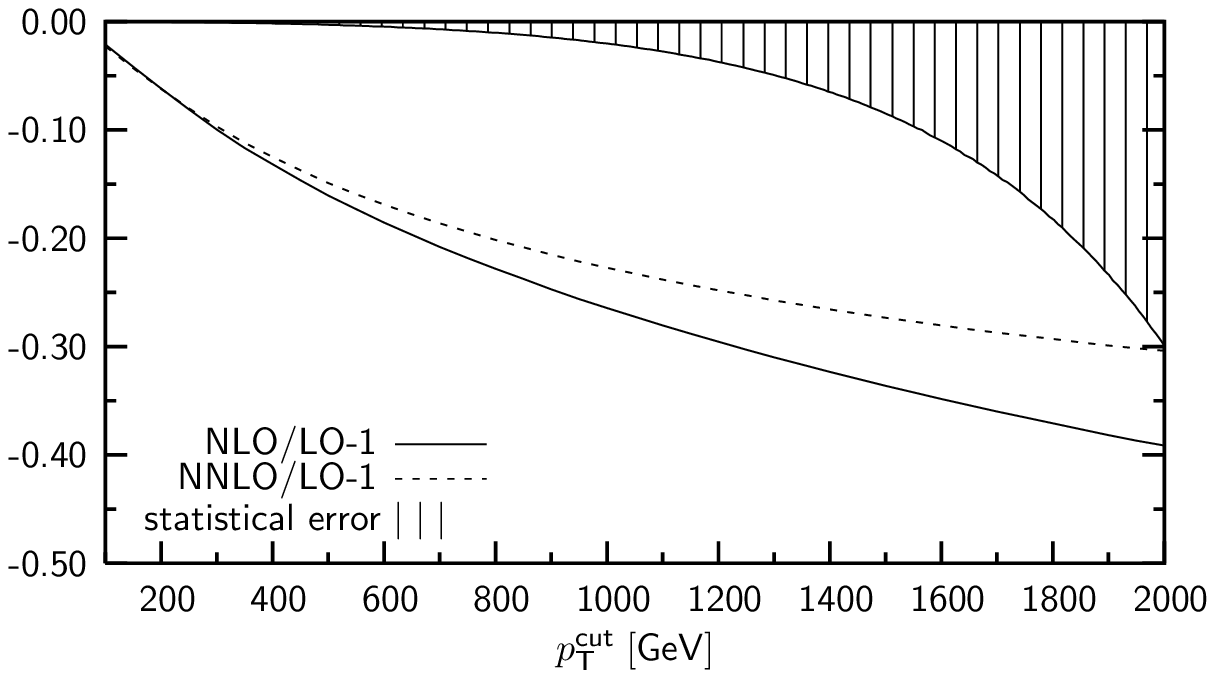, angle=0, width=11.5cm}
\end{center}
\vspace*{-2mm}
\caption{
Relative NLO (solid) and NNLO (dotted) corrections 
\wrt the LO prediction  and statistical
  error (shaded area) for the unpolarized integrated cross section for $pp\rar Z j$ at $\sqrt{s}=14 \TeV$ as a function of $\pTcut$.
}
\label{fig:lhctot}
\end{figure}

To underline the relevance of these effects, 
in \reffi{fig:lhctot} we present 
the relative NLO and NNLO 
corrections for the cross section,
integrated over $\pT$ starting from $\pT = \pTcut$, as a function of
$\pTcut$. 
This is compared with the statistical error, defined as
$\Delta \si_{\rm stat} / \si = 1 /\sqrt N$ with $N= \calL \times {\rm BR}
(Z\rar l,\nu_l)\times \si_{\rm LO}$. We include all leptonic decay
modes of $Z$, 
corresponding to  ${\rm BR}=30.6 \%$, and assume a total integrated
luminosity $\calL =300 \fba^{-1}$ for the LHC~\cite{LHClum}. It is clear from
\reffi{fig:lhctot}, that the size of the one- and two-loop corrections 
is much bigger than and comparable to the statistical error, respectively.

\begin{figure}[]
\vspace*{2mm}
  \begin{center}
\epsfig{file=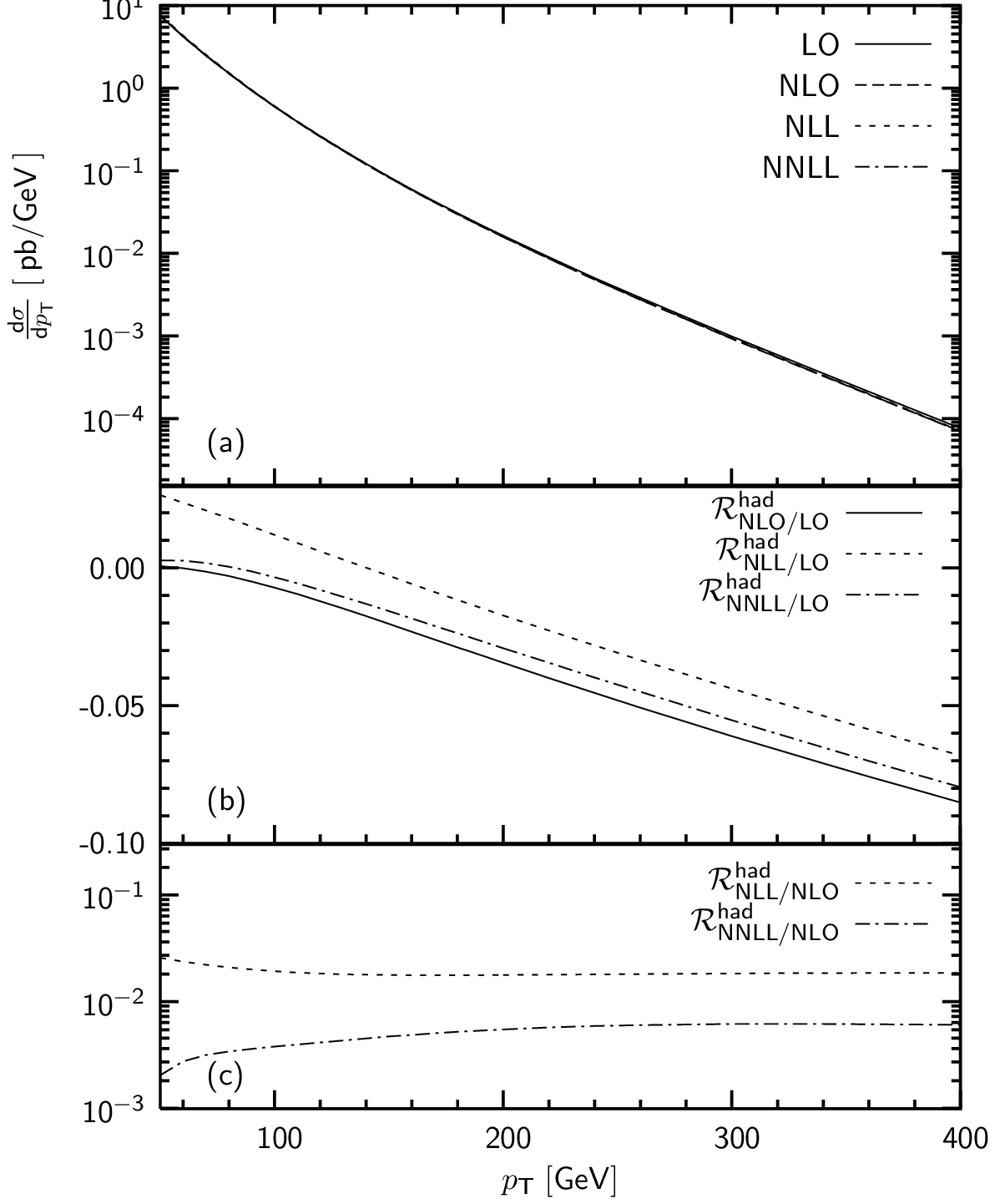, angle=0, width=11.5cm}
\end{center}
\vspace*{-2mm}
\caption{Transverse momentum distribution for $p\bar p\rar Z j$ at
  $\sqrt{s}=2 \TeV$.
(a) LO (solid), NLO (dashed),  NLL (dotted) and 
NNLL (dot-dashed) predictions. 
(b) Relative NLO (solid), NLL (dotted) and NNLL (dot-dashed)
weak correction \wrt the LO distribution.
(c) NLL (dotted) and NNLL (dot-dashed) approximations compared to the 
full NLO result.
}
\label{fig:tev}
\end{figure}

\begin{figure}[]
\vspace*{2mm}
  \begin{center}
\epsfig{file=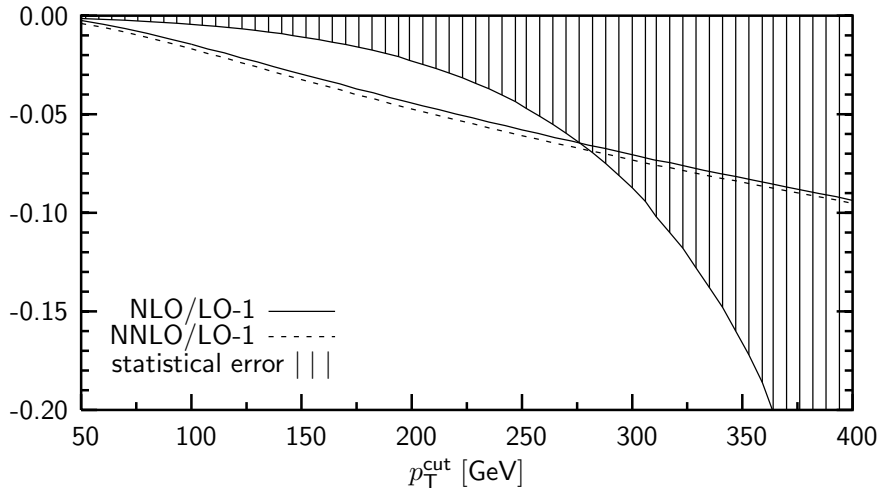, angle=0, width=11.5cm}
\end{center}
\vspace*{-2mm}
\caption{Relative NLO (solid) and NNLO (dotted) corrections \wrt the LO and statistical
  error (shaded area) for the unpolarized integrated cross section for $p\bar p\rar
  Z j$ at $\sqrt{s}= 2 \TeV$ as a function of $\pTcut$.
}
\label{fig:tevtot}
\end{figure}

We also perform a similar analysis for
high transverse momentum $Z$ production at the Tevatron.
In  \reffi{fig:tev} we show the transverse momentum distribution
(\reffi{fig:tev}a), the relative size of the corrections (\reffi{fig:tev}b)
and the quality of the one-loop NLL and NNLL approximations (\reffi{fig:tev}c).
In \reffi{fig:tevtot} the relative size of the corrections 
to the integrated cross section is compared 
with the statistical error expected for an integrated 
luminosity  $\calL =11 \fba^{-1}$~\cite{TEVlum}.
At the energies of the Fermilab collider,
the NLO weak correction is of the order of the 
statistical error and should be taken into account 
when considering precision measurements.
The two-loop terms turn out to be negligible.
 
\begin{figure}[]
\vspace*{2mm}
  \begin{center}
\epsfig{file=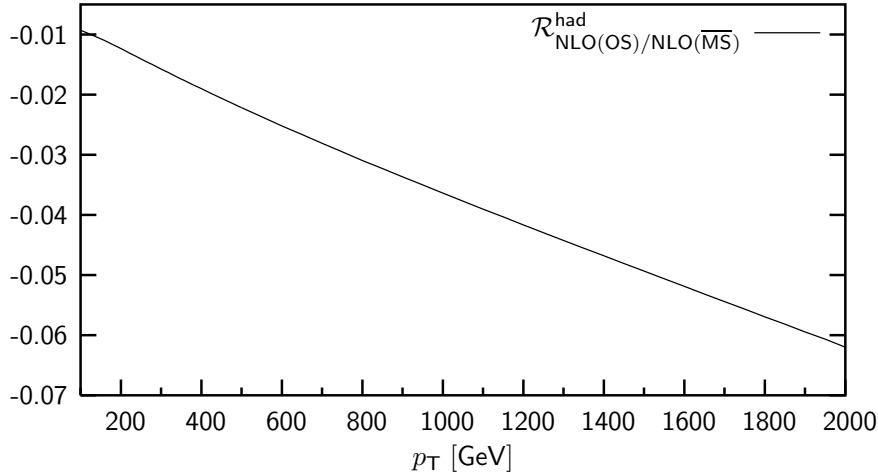, angle=0, width=11.5cm}
\end{center}
\vspace*{-2mm}
\caption{Relative difference of the NLO transverse momentum distribution
for $pp\rar Z j$ calculated in  the \msbar~and on-shell (OS) schemes 
at $\sqrt{s}=14 \TeV$.
}
\label{fig:schemes}
\end{figure}

Finally, in \reffi{fig:schemes} we illustrate the dependence of the NLO $\pT$
distribution on the choice of the renormalization scheme. 
At low
$\pT$ the results in the on-shell
and \msbar~schemes differ by around 1\%
and the difference grows with $\pT$ reaching 6\% at $\pT=2 \TeV$. 
This effect is mainly due to the different
treatment of the weak mixing angle in the 
two renormalization schemes.
As well known, the relation between 
the \msbar~and on-shell 
definitions of the weak mixing angle 
is provided by the $\rho$ parameter as \cite{Degrassi:1990tu}
\beq
\frac{\cw^2}{\hat\cw^2}
=
\frac{M_W^2}{M_Z^2 \,\hat\cw^2}
=
\rho
,\qquad
\frac{\hat\sw^2}{\sw^2}
-1=
\Delta\sw^{2}=
\frac{\cw^2}{\sw^2}
\Delta\rho
,
\eeq
where $\rho=(1-\Delta \rho)^{-1}$
and the symbols with and without hat denote \msbar~and 
on-shell quantities, respectively.
The input parameters used in our calculation,
$\hat\sw^2=0.2314$ and $\sw^2=1-M_W^2/M_Z^2\simeq 0.2228$,
correspond to $\Delta \sw^2\simeq 3.8\%$ and are
extracted from precision electroweak measurements
taking all available loop corrections into account. 
Instead, loop corrections beyond $\ord(\alpha)$ are not 
included in our calculation and, in particular, 
the deviation observed in \reffi{fig:schemes}
is due to missing two-loop (and higher-order) corrections 
related to the $\rho$ parameter.
The scheme dependence resulting from $\alpha/\sw^2$ terms 
amounts to
\beq\label{schemdep1}
\frac{\alpha}{\sw^2}
({1-\Delta^{(1)}\sw^2})
-
\frac{\alpha}{\hat\sw^2}
\simeq
\frac{\alpha}{\hat\sw^2}
(\Delta \sw^2-\Delta^{(1)}\sw^2),
\eeq
where $\Delta^{(1)}\sw^2=(\cw^2/\sw^2)\Delta \rho^{(1)}\simeq 4.5 \%$
corresponds to the one-loop corrections to the $\rho$ parameter,
which are included in our on-shell predictions through the counterterm
 \refeq{weakmixct}.
This scheme dependence \refeq{schemdep1} is thus due to the higher-order contributions
$\Delta\sw^2 - \Delta^{(1)}\sw^2\simeq -0.7\%$.
Their relatively large size results from the combined effect of
 $\ord(\alpha\alpha_\rS  m_t^2)$
\cite{Djouadi:1987gn},
$\ord( \alpha\alpha_\rS^2 m_t^2)$
\cite{Chetyrkin:1995ix},
$\ord(\alpha^2 m_t^4)$
\cite{Barbieri:1992nz}
and
$\ord(\alpha^2 m_t^2)$
\cite{Degrassi:1996mg}
corrections to the $\rho$ parameter 
and is consistent with the effect observed in 
 \reffi{fig:schemes} at small $p_\rT$.
In addition,
the scheme-dependence resulting from 
the one-loop logarithmic terms is of order
\beq\label{schemdep2}
-\left(\frac{\alpha^2}{\sw^4}
-
\frac{\alpha^2}{\hat\sw^4}\right)
\log^2(\shat/M_W^2)
\simeq
-2\Delta\sw^2
\frac{\alpha^2}{\sw^4}
\log^2(\shat/M_W^2).
\eeq 
This effect is due 
to missing two-loop corrections of order 
$\Delta \rho \;\alpha\log^2(\shat/M_W^2)$.
Its size is proportional to $\Delta \sw^2\simeq 3.8\%$ 
and grows with energy. 
This explains the high-$p_\rT$ behaviour in \reffi{fig:schemes}.

We stress that the effects 
\refeq{schemdep1}--\refeq{schemdep2}
are entirely due to missing higher-order terms related to $\Delta \rho$
and that such missing terms concern only the calculation in the on-shell scheme.
Indeed, $\Delta \rho$ enters our predictions only through the relation between 
the weak mixing angle and the weak-boson masses in the on-shell scheme
whereas the \msbar~calculation does not receive any contribution from $\Delta \rho$.
The large scheme-dependence in \reffi{fig:schemes}
has thus to be interpreted as large uncertainty of the one-loop prediction 
in the on-shell scheme whereas such uncertainties are absent in the  \msbar~scheme.
This motivates the choice of the  \msbar~scheme
adopted in this paper.

\section{Summary}
\label{se:conc}
In this work we have calculated the one-loop weak corrections to hadronic 
production of $Z$ bosons at large 
transverse momenta.
Analytical results are presented for the parton subprocess 
$\bar q q \to Z g$ and its crossed versions.
Special attention has been devoted to the high-energy region, $\shat\gg M_W^2$, 
where the weak corrections are enhanced by logarithms of $\shat/M_W^2$. 
For this region we have derived approximate expressions that include 
all large logarithms  as well as those terms that are not logarithmically 
enhanced. 
The quadratic and linear logarithms confirm earlier results obtained to  
next-to-leading logarithmic accuracy. 
The complete high-energy approximation presented in this paper is in 
excellent agreement with the exact one-loop predictions.
We give numerical results for 
proton-antiproton collisions at 2 TeV (Tevatron)
and proton-proton collisions at 14 TeV (LHC)
in the region of large tranverse momentum  ($p_\rT$).
The corrections are negative and their size increases with $p_\rT$.
At the Tevatron, transverse momenta up to 300 GeV will be explored 
and the weak corrections 
may reach up to $-7\%$.
At the LHC, transverse momenta of 2 TeV or more are within the reach.
In this region the 
corrections are large,
$-30\%$ up to $-40\%$, 
and even the dominant two-loop 
logarithmic terms must be included in any realistic prediction.

\begin{appendix}

\section{Analytical results}
\label{app:coeff}

Here we present the explicit analytical expressions
for the coefficients 
$K_j^{\mathrm{A/N}}(M_V^2)$,
which are defined in \refeq{implcoeff} and 
correspond to the Abelian (A) and non-Abelian (N) contributions 
resulting from the Feynman diagrams of \reffi{fig:loopdiags}.

\subsubsection*{Abelian coefficients}
\beqar\label{nloabcoeff}
K^{\mathrm{A}}_{0 }(M_V^2) 
&=& 
-\frac{4 \shat^2+3(\that^2+\uhat^2)}{\that \uhat}
+\shat\Bigg(\frac{1}{\shat+\that}+\frac{1}{\shat+\uhat}-\frac{5}{\uhat}-\frac{5}{\that}+\frac{4}{\that+\uhat}\Bigg)
,\nl
K^{\mathrm{A}}_{1 }(M_V^2) 
&=&  
-\Bigg[\frac{3 \shat}{{{(\shat+\that)}^2}}+\frac{3 \shat}{{{(\shat+\uhat)}^2}}\Bigg]
+
\Bigg(\frac{1}{\shat+\that}+\frac{1}{\shat+\uhat}\Bigg)
-2\Bigg(\frac{ \shat+\uhat}{{\that^2}}+\frac{ \shat+\that}{{\uhat^2}}\Bigg)
\nl&&{}
+\frac{ 2 {\shat^2} (2\shat+\that+\uhat)}{ \that \uhat (\shat+\that)  (\shat+\uhat)}
+4\frac{ (\shat+\that)^2+(\shat+\uhat)^2}{ \that \uhat M_V^2}
,\nl
K^{\mathrm{A}}_{2 }(M_V^2) 
&=& 
M_Z^2 \Bigg[ \frac{6 \shat M_V^2}{{{(\shat+\that)}^3}}+\frac{6 \shat M_V^2}{{{(\shat+\uhat)}^3}}+\frac{2 \shat M_V^2}{{{(\shat+\that)}^2} \uhat}+\frac{2 \shat M_V^2}{
{{(\shat+\uhat)}^2\that}}+\frac{4 (\shat+\that+\uhat)}{{{(\that+\uhat)}^2}}
-\frac{3}{\that}
\nl&&{}
-\frac{3}{\uhat}
+\frac{2 \shat+\that-2 M_V^2}{{{(\shat+\that)}^2}}
+\frac{2 \shat+\uhat-2 M_V^2}{{{(\shat+\uhat)}^2}}-\frac{\shat
(2\shat+\that+\uhat) (2 M_V^2+3 \shat)}{\that \uhat (\shat+\that) (\shat+\uhat)}\Bigg]
,\nl
K^{\mathrm{A}}_{3 }(M_V^2) 
&=&  
0
,\nl
K^{\mathrm{A}}_{4 }(M_V^2) 
&=& 
-\frac{4 \shat (\shat+2\that+2\uhat)}{{{(\that+\uhat)}^2}}
,\nl
K^{\mathrm{A}}_{5 }(M_V^2) 
&=& 
-\frac{6 M_V^2 \shat \uhat }{{{(\shat+\that)}^3}}
+\frac{M_V^2 (2 \uhat-5 \shat)-\shat
\uhat}{{{(\shat+\that)}^2}}
+\frac{2 M_V^2 (\shat+\that+\uhat) }{{\uhat^2}}-\frac{M_V^2+4 \shat+\uhat}{\shat+\that}
,\nl
K^{\mathrm{A}}_{6 }(M_V^2) 
&=& K^{\mathrm{A}}_{5 }(M_V^2)\Big|_{\that\leftrightarrow\uhat} 
,\nl
K^{\mathrm{A}}_{7 }(M_V^2) 
&=&
-\frac{\shat }{\that \uhat} \Bigg[2 (\shat+M_V^2)(\that+\uhat) +{\that^2}+{\uhat^2}\Bigg]
,\nl
K^{\mathrm{A}}_{8 }(M_V^2) 
&=&
\frac{M_Z^2 M_V^2}{\uhat(\uhat-M_Z^2)^3}
\Bigg[
2\that M_V^2 (\uhat-\shat-\that)
-4 M_Z^2\shat (\shat+\that+M_V^2)
\Bigg]
,\nl
K^{\mathrm{A}}_{9 }(M_V^2) 
&=&  
0 
,\nl
K^{\mathrm{A}}_{10}(M_V^2) 
&=& K^{\mathrm{A}}_{8}(M_V^2)\Big|_{\that\leftrightarrow\uhat} 
,\nl
K^{\mathrm{A}}_{11}(M_V^2) 
&=&  
0
,\nl
K^{\mathrm{A}}_{12}(M_V^2) 
&=& 
-\frac{M_V^2(\that+\uhat)+\shat\uhat
}{\that \uhat}
\Bigg[2{{(\shat+M_V^2)}}(\shat+M_V^2+\that)+{\that^2}\Bigg] 
,\nl
K^{\mathrm{A}}_{13}(M_V^2) 
&=& K^{\mathrm{A}}_{12}(M_V^2) \Big|_{\that\leftrightarrow\uhat} 
,\nl
K^{\mathrm{A}}_{14}(M_V^2) 
&=& 
0
.
\eeqar

\subsubsection*{Non-Abelian coefficients}

\beqar\label{nlonabcoeff}
K^{\mathrm{N}}_{0 }(M_W^2) 
&=& 
\frac{4 {\shat}}{\that \uhat}(\shat+\that+\uhat)
-2 \shat \Bigg(\frac{1}{\shat+\uhat}+\frac{1}{\shat+\that}+\frac{2}{\that+\uhat}\Bigg)+2
\Bigg(\frac{\that}{\uhat}+\frac{\uhat}{\that}\Bigg)
,\nl
K^{\mathrm{N}}_{1}(M_W^2) 
&=&  
0
,\nl
K^{\mathrm{N}}_{2}(M_W^2) 
&=& 
-K^{\mathrm{A}}_{2}(M_W^2) 
,\nl
K^{\mathrm{N}}_{3}(M_W^2) 
&=& 2 {M_W^2} \Bigg[\frac{3 \shat \that}{{{(\shat+\uhat)}^3}}+\frac{3 \shat \uhat}{{{(\shat+\that)}^3}}\Bigg]
-\frac{1}{\that\uhat}
\Bigg[ \frac{1}{(\shat+\that)^2}+ \frac{1}{(\shat+\uhat)^2}\Bigg]
\Bigg\{
{\shat^4}
-2 {\that^2} {\uhat^2}
\nl&&{}
+\shat^2(\that+\uhat)( 2\shat+\that+\uhat )
+2 {M_W^2} \Bigg[{\shat^2}(\shat+\that+\uhat) 
- \that\uhat( 2 \shat-\that -\uhat )\Bigg]
\Bigg\}
,\nl
K^{\mathrm{N}}_{4}(M_W^2) 
&=& 
-K^{\mathrm{A}}_{4}(M_W^2) 
,\nl
K^{\mathrm{N}}_{5}(M_W^2) 
&=& 
\frac{2 \shat (\shat+\that)-2 \that \uhat}{{{(\shat+\that)}^2}}
,\nl
K^{\mathrm{N}}_{6}(M_W^2) 
&=& K^{\mathrm{N}}_{5 }(M_W^2)\Big|_{\that\leftrightarrow\uhat} 
,\nl
K^{\mathrm{N}}_{7}(M_W^2) 
&= &
-K^{\mathrm{A}}_{7}(M_W^2) 
,\nl
K^{\mathrm{N}}_{8}(M_W^2) 
&=& 
-K^{\mathrm{A}}_{8}(M_W^2)
,\nl
K^{\mathrm{N}}_{9}(M_W^2) 
&=& 
2 \uhat-\frac{\shat^2-\shat\that}{\that}+\frac{{\shat^2}+ \shat\that}{\uhat}
+2{M_W^2} \Bigg[\frac{2 {\shat^2}+\that \shat+{\that^2}}{\that \uhat}-\frac{2 \that \uhat}{{{(\shat+\that)}^2}}-\frac{\that-\shat}{\that}\Bigg] 
\nl&&{}
+\frac{2  {M_W^4} }{\shat+\that}
\Bigg[\frac{{\shat^2}}{\that \uhat}-\frac{\uhat(2\shat-\that)}{{(\shat+\that)}^2}-\frac{4 \shat}{\shat+\that}\Bigg]
,\nl
K^{\mathrm{N}}_{10}(M_W^2) 
&=&  K^{\mathrm{N}}_{8 }(M_W^2)\Big|_{\that\leftrightarrow\uhat} 
=-K^{\mathrm{A}}_{10}(M_W^2) 
,\nl
K^{\mathrm{N}}_{11}(M_W^2) 
&=& 
 K^{\mathrm{N}}_{9 }(M_W^2)\Big|_{\that\leftrightarrow\uhat}
,\nl
K^{\mathrm{N}}_{12}(M_W^2) 
&=& 
-K^{\mathrm{A}}_{12}(M_W^2) 
,\nl
K^{\mathrm{N}}_{13}(M_W^2) 
&=& K^{\mathrm{N}}_{12}(M_W^2)\Big|_{\that\leftrightarrow\uhat} 
=-K^{\mathrm{A}}_{13}(M_W^2) 
,\nl
K^{\mathrm{N}}_{14}(M_W^2) 
&=& 
\frac{{M_W^2} (\that+\uhat)-\that \uhat}{\that \uhat} 
\Bigg[2 {M_W^4}+2 (2 \shat+\that+\uhat) {M_W^2}-2
\that \uhat-\shat (\that+\uhat)\Bigg]
.
\eeqar

\end{appendix}

\section*{Acknowledgements}

S.~P.~is grateful to A.~Denner for providing a set of 
routines for the numerical evaluation of one-loop integrals.
This work was supported in part by 
BMBF Grant No.~05HT4VKA/3
and by the Deutsche Forschungsgemeinschaft 
(Sonderforschungsbereich Transregio SFB/TR-9 
``Computational Particle Physics''). 
M.~S. would like to acknowledge financial 
support from the Graduiertenkolleg "Hochenergiephysik und Teilchenastrophysik".

\end{document}